\documentclass[
prd,
superscriptaddress,
10 pt,
preprintnumbers,
notitlepage,
secnumarabic,
nobibnotes,
nofootinbib,
showpacs
]{revtex4-1}

\usepackage[T1]{fontenc}
\usepackage[pdftex]{color,graphicx}
\usepackage{mathtools}
\usepackage{amssymb}
\usepackage{bm}
\usepackage{dsfont}
\usepackage{mathrsfs}
\usepackage{calligra}
\usepackage{tabularx}
\usepackage[artemisia]{textgreek}
\usepackage{hyperref}
\usepackage{xcolor}
\usepackage{enumerate}
\usepackage[normalem]{ulem}
\usepackage{multirow}
\usepackage{caption}
\usepackage{perpage}
\MakePerPage{footnote}
\renewcommand{\.}{\!\;}		
\renewcommand{\@}{\!\:\!}	
\newcommand\atopp[2]{\genfrac{}{}{0pt}{}{#1}{#2}}
\DeclareMathAlphabet\mathbfcal{OMS}{cmsy}{b}{n}
\newcommand{\kbar}{\mathchar'26\mkern-9mu k}
\renewcommand{\i}{\!\:\mathrm{i}}
\newcommand{\und}[1]{\underline{#1}}

\newcommand{\cev}[1]{\reflectbox{\ensuremath{\vec{\reflectbox{\ensuremath{#1}}}}}}

\usepackage{tikz}
\tikzset{c/.style={draw=black, line width=1pt}}
\tikzset{d/.style={dashed, draw=lightgray, line width=0.5pt}}
\tikzset{l/.style={draw=black, line width=1.5pt}}
\tikzset{ll/.style={draw=black, line width=1pt}}
\tikzset{lll/.style={draw=black, line width=0.5pt}}

\begin{document}
\title{Regularization of the cosmological sector of loop quantum gravity with bosonic matter
\\
and the related problems with general covariance of quantum corrections}

\author{Jakub Bilski}
\email{bilski@zjut.edu.cn}
\affiliation{Institute for Theoretical Physics and Cosmology, Zhejiang University of Technology, 310023 Hangzhou, China}


\begin{abstract}
\noindent
This article concerns the problems regarding different lattice regularization techniques for the matter fields of Hamiltonian constraints defined in the framework of loop quantum gravity. The analysis is formulated in the phase space-reduced cosmological model of the hypothetical theory of canonical quantum general relativity. This article explains why a different than links-related lattice smearing of fields leads to a local violation of general covariance. This happens by assuming, for instance, the nodes-related smearing. Therefore, this problem occurs in the case of any polymerlike scalar field quantization method by breaking the background independence of the semiclassical predictions. In consequence, the diffeomorphism symmetry that depends on a links distribution is broken locally at the level of generally relativistic corrections. Moreover, by using the phase space-reduced gauge fixing technique to analyze this issue, the results are general and they concern any coupling with the links-regularized gravitational degrees of freedom in loop quantum gravity. Therefore, they lead to the following no-go conclusion. Any lattice smearing of matter, not defined by using the geometrical distribution specified by the links-fluxes duality, violates the general principle of relativity.
\end{abstract}

\maketitle


\section{Introduction}\label{I}

\noindent
Nonperturbative quantum gravity is a theoretical branch of physics, which assumes that the unification of the principles of general relativity (GR) with a possibly quantum nature of the gravitational field is probable. The quantum theory that aims to capture the restrictions of the strong equivalence principle is loop quantum gravity (LQG) \cite{Thiemann:1996aw,Thiemann:2007zz}. In this model, the gravitational field is described by the Ashtekar variables \cite{Ashtekar:1986yd}. Its kinematical structure is similar to one in the SU$(2)$ Yang-Mills field \cite{Yang:1954ek}. By following the example of the regularization in quantum chromodynamics \cite{Wilson:1974sk}, LQG is formulated in terms of $\mathfrak{su}(2)$-invariant variables on a lattice \cite{Ashtekar:1986yd}. This theory, however, does not lead to the generally covariant description of experiments and observations. The general postulate of relativity \cite{Einstein:1916vd} in LQG concerns only the gravitational field, which is described equivalently in all coordinate systems. However, a procedure providing measurable predictions, which should not depend on any character of an observer's frame, is not uniquely determined even only for this field. This equivalent description in LQG is known as the strong formulation of systems equivalence (SE) for gravitation (the equivalence principle without its standard Einstein's version for matter fields) \cite{Einstein:1907,Einstein:1911vc}. The observer-independent predictions are known as the background independence (BI) of the related observations (the law of general covariance) \cite{Einstein:1907iag,Einstein:1916vd}.

The fundamental consistency of the theory requires a unique procedure of the quantum corrections derivation in LQG, which will guarantee the BI of this framework. Almost any research toward possibly detectable predictions of quantum gravity requires also an extension of SE and BI to matter. Only pure gravitational experiments or observations would not need this extension. Therefore, physical studies on gravitational waves, which involve their possible quantum nature could be described by LQG. However, any cosmological research of the interactions between quantum spacetime and matter requires a more general framework.

The first unified construction of GR and quantum field theories of matter interactions (QFT), formulated as an extension of LQG, is known as canonical quantum general relativity (CQGR) \cite{Thiemann:2007zz}. In this article, CQGR is going to have a more general meaning. We use this name for any hypothetical theory, which generalizes the interactions between the Newtonian gravitational field and QFT into a consistent quantum model that satisfies SE and BI, and which is quantized in the canonical procedure \cite{DeWitt:1967yk}.

Although, any complete formulation of CQGR does not exist, one can investigate a coupling between LQG and a simple model of a bosonic matter field that has SE and BI. By assuming that both the gravitational and matter fields satisfy the general postulates of relativity and by constructing the quantization of these fields which preserves these postulates, one can formulate the following hypothesis. The postulated model can correctly describe physical interactions between gravity and bosonic matter by satisfying the physical constructional requirements. Therefore, it could be worth it to study the phenomenological predictions also of the toy model that would be a symmetry-preserving simplification of a more general theory. However, by finding agreements between the predictions and the related measurements, one could not claim the truthfulness or universality of this model yet. This would only increase the probability that the toy model is the simplified version of a correct fundamental theory. In consequence, it would be worth it to look for a theory, the particular limit of which is this toy model that was found to be in agreement with the data.

Conversely, one can also construct an inconsistent toy model in purpose, for instance, because in this way it would be simpler. The result, however, could be used only for theoretical analysis; for instance to verify the behavior of a particular mathematical procedure. Using this inconsistent model to study phenomenology would not have any physical value. Moreover, if one would do it anyhow and find any agreement between the predictions and data, this would tell nothing about the nature of the only apparently `predicted' phenomenon.

By following the preceding methodology of the qualitative evaluation of toy models, several proposed cosmological theories based on the framework of LQG are going to be tested as the potential candidates to study phenomenology. This directly involves the consistency verification of the restrictions and methods regarding whether the formulation of quantum gravitational fields and quantum matter fields is SE and BI. Both types of these fields must satisfy both conditions and be independently quantizable.

To demonstrate that a model is inconsistent is enough to indicate a single violation of the methodological assumptions. Hence, the investigation of the less strict and recurrent element of quantum cosmological models is going to be studied. To find problems regarding this element is more probable. Therefore, the matter sector of different approaches to quantum cosmology and the BI of its semiclassical limit is analyzed in this article. The general investigation of this issue would require detailed studies for each model separately. To avoid this complication, a maximal simplification of these theories is going to be assumed. In what follows it will be enough to consider the formalism of anisotropic loop quantum cosmology (LQC) \cite{Ashtekar:2009vc,Ashtekar:2011ni}. The SE gravitational degrees of freedom description is going to be coupled with the SE framework of the scalar field \cite{Thiemann:1997rt,Bilski:2015dra}. This is the simplest system, which could be considered as a cosmological phase space reduction \cite{Bilski:2019tji} of CQGR. As the reader will see, the violation of the general covariance in this system will be found. Therefore, it is worth to introduce one more reference matter field, the quantization of which is more similar to the one of gravity. This is the vector field \cite{Thiemann:1997rt,Bilski:2016pib}, and its simplest Abelian version will be considered. Finally, the formalism linking the isotropic and anisotropic cosmology with the vector matter field will be derived to demonstrate how BI, which is violated in the scalar field case, can be preserved in the theory. The essential part of this analysis is going to be the investigation of the so-called inverse volume corrections in LQG.
\\

In this article the standard framework of canonical LQG is considered. The $3\@+\@1$ decomposition of a manifold $M$ that represents spacetime is introduced by the foliation into Cauchy hypersurfaces $\Sigma_t$ \cite{Arnowitt:1960es,Arnowitt:1962hi}. The tetrad formalism with an internal SU$(2)$ symmetry is applied and the time gauge is assumed. In this article, the gravitational coupling constant is defined as $\kappa=16\pi G$, where the speed of light is normalized to $c=1$. The fundamental constant for the canonical DeWitt quantization \cite{DeWitt:1967yk} is defined as $\kbar:=\frac{1}{2}\gamma\hbar\kappa=8\pi\gamma l_P^2$, where $\gamma$ and $l_P$ are the real Immirzi parameter and the Planck length, respectively. The repeated indices written in $(\ )$ brackets are not summed; all the other indices follow the Einstein summation convention.

The article is organized as follows. In Sec.~\ref{II} the lattice regularization of bosonic fields is introduced. In Sec.~\ref{III} the phase space-reduced cosmological framework of CQGR is defined. Then the verification of general covariance is done in Sec.~\ref{IV}. The conclusions of the article are that the node-related regularization of the matter sector leads to background-dependent predictions. The general postulate of relativity can be preserved by considering only the links smearing of all the propagating degrees of freedom in CQGR.


\section{Regularization}\label{II}

\subsection{Lattice Yang-Mills theory}\label{II.1}

\noindent
Two examples of the Yang-Mills field \cite{Yang:1954ek} are important concerning the cosmological analysis in this article. The simplest representative of the matter vector field is the Abelian Yang-Mills field. Its non-Abelian variant that satisfies the $\mathfrak{su}(2)$ algebra describes gravity \cite{Ashtekar:1986yd}.

Let $l:[0,1]\to\Sigma_t$ be a smooth path parametrized by $s\in[0,1]$ and located inside the constant time surface $\Sigma_t$ constructed by the Arnowitt-Deser-Misner (ADM) method \cite{Arnowitt:1960es,Arnowitt:1962hi}. One can define an embedding of $l(s)$ in $\Sigma_t$ and introduce a parameter $\varepsilon$ such that $l_{\varepsilon}(s):=l(\varepsilon s)$. In this article $\varepsilon$ has the dimension of a length, its maximal value is restricted by the subsequent definition and the minimal one by the choice of the so-called shadow states \cite{Ashtekar:2002sn} (the coherent states in LQC); this leads to the inequality $1>\varepsilon^2>|\gamma|\.l_P^2$ that is implicitly expressed in some length scale units\footnote{This assumption on the one hand decreases the universality of this analysis, but on the other hand, allows to quickly compare the obtained results with the most popular LQC's framework-related models. It is worth noting that an improved approach to the lattice regularization \cite{Bilski:2020poi} and the related cosmologically reduced model \cite{Bilski:2021_LCC} would lead to the same consclusions concerning the structure of the semiclassical corrections. In this case the lower cutoff on $\varepsilon^2$ would not be needed.}. The parallel transport equation for a vector $u(s)$ along $l_{\varepsilon}(s)$ reads
\begin{align}
\label{parallel}
\und{\partial}_{\dot{l}_{\varepsilon}}\und{u}(s)=\frac{d}{ds}\und{u}(s)+\und{A}\big(\dot{l}_{\varepsilon}(s)\big)\und{u}(s)=0\,.
\end{align}
It has the following solution: $\und{u}(s)=\big(\und{h}_{l_{\varepsilon}(s)}\big)^{\!-1}\und{u}(0)$, known as a holonomy, where
\begin{align}
\label{holonomy}
\und{h}_{l_{\varepsilon}}:=\mathcal{P}\exp\!\bigg(\!\int_0^{1}\!\!\!\!ds\,\und{A}\big(\dot{l}_{\varepsilon}(s)\big)\!\bigg).
\end{align}

The propagating degrees of freedom of the Abelian matter vector field $\und{A}_{\mu}$ are introduced by the action
\begin{align}\label{YMaction}
S^{(\und{A})}:=
-\frac{1}{4\mathsf{g}_{\!\und{A}}^2}\!\int_{\!M}\!\!\!d^{4}x\sqrt{-g}\,g^{\mu\nu}g^{\xi\pi}\und{F}_{\mu\xi}\und{F}_{\nu\pi}\,,
\end{align}
where $g_{\mu\nu}$, $g^{\mu\nu}$, and $g$ are the metric tensor, its inverse, and determinant, respectively. The coupling constant is denoted by $\mathsf{g}_{\!\und{A}}^2$ and $\und{F}_{\mu\xi}$ is the curvature of $\und{A}_{\mu}$.

The $3+1$ spacetime splitting allows one to derive the momentum $\und{E}^a=\frac{\sqrt{q}}{\mathsf{g}_{\!\und{A}}^2}e_0^{\mu}q^{ab}\und{F}_{\mu b}$ canonically conjugated to $\und{A}_{a}$. Here, $q$ denotes the determinant of the $q_{ab}:=e^i_ae^i_b$ metric on $\Sigma_t$ and $e_0^{\mu}=(1/N,-N^a/N)$ is the upper row of the vierbein matrix $e_{\alpha}^{\mu}$ (`$\alpha$' represents directions in the Minkowski space).

The Legendre transform of \eqref{YMaction} leads to the completely constrained system with the total Hamiltonian $H^{(\und{A})}_T=V^{(\und{A})}+H^{(\und{A})}$ that is composed of two first-class constraints. The vector constraint (called also the diffeomorphism constraint)
\begin{align}\label{YMvector}
V^{(\und{A})}:=\int_{\Sigma_t}\!\!\!\!d^3x\,N^a\mathcal{V}_a^{(\und{A})}
=\int_{\Sigma_t}\!\!\!\!d^3x\,N^a\und{F}_{ab}\und{E}^b
\end{align}
imposes the invariance under the spatial diffeomorphism transformations. The Hamiltonian constraint (called also the scalar constraint)
\begin{align}\label{YMscalar}
H^{(\und{A})}:=\int_{\Sigma_t}\!\!\!\!d^3x\,N\mathcal{H}^{(\und{A})}
=\frac{\mathsf{g}_{\!\und{A}}^2\!}{2}\!\int_{\Sigma_t}\!\!\!\!d^3x\,N\frac{1}{\sqrt{q}}q_{ab}
\big(\und{E}^a\und{E}^b+\und{B}^a\und{B}^b\big)
\end{align}
generates the time reparametrization symmetry. The last quantity in the preceding equation is the magnetic field $\und{B}^a:=\frac{1}{2\mathsf{g}_{\!\und{A}}^2\!}\epsilon^{abc}\und{F}_{bc}$, where $\epsilon^{abc}:=\sqrt{q}\tilde{\epsilon}^{abc}$ and $\tilde{\epsilon}^{abc}$ is the Levi-Civita tensor. It is worth noting that $\und{E}^a$ and $\und{B}^a$ are a vector density and a pseudovector density, respectively. By being densities, these objects scale properly according to the scaling of the integration measure $d^3x$, where $d^3x\sqrt{q}$ is the measure invariant in $\mathds{R}^3$ and $\sqrt{q}$ is a weight-$1$ scalar density.

In the lattice framework the vector constraint in \eqref{YMvector} is added to its gravitational analog and they are solved at the classical level. The Hamiltonian constraint in \eqref{YMscalar} is quantized after the regularization of the canonical fields on the diffeomorphisms-invariant lattice. This leads to the construction of the Hamiltonian constraint operator (HCO), which is the only element of $H^{(\und{A})}_T$ that is going to be solved at the quantum level.

The regularization procedure assumes the introduction of the Wilson loops \cite{Wilson:1974sk}. In the Abelian case they trivially reduce to loop holonomies
\begin{align}
\label{YMWilson}
\und{h}_{l\circlearrowleft l'}
=\varepsilon_l\varepsilon_{l'}\und{F}_{ab}\dot{l}^a\dot{l}'^b+\mathcal{O}(\varepsilon^3)\,,
\end{align}
where the loop begins at the initial point of the $l$ link, goes along a quadrilateral path (in the cosmological framework in this article) and returns to the same point along $l'$. The second lattice-regularized variable takes the form of the $\und{E}^a=\frac{1}{2}\epsilon^{abc}{\,}^*\!\und{E}_{bc}$ field flux, constructed by smearing the two-form ${\,}^*\!\und{E}_{bc}$ (Hodge dual to $\und{E}^a$) over a two-dimensional surface $\mathbf{S}$,
\begin{align}
\label{YMflux}
\und{f}(\mathbf{S}):=\int_{\mathbf{S}}\!^*\!\und{E}=\int_{\mathbf{S}}\!n_a\und{E}^{a}\,,
\end{align}
where $n_a:=\epsilon_{abc}dx^b\wedge dx^c$ is the normal to $\mathbf{S}$.

\subsection{GR in terms of Ashtekar variables}\label{II.3}

\noindent
The gravitational degrees of freedom are represented by the non-Abelian real vector field $A^i_a:=\frac{1}{2}\epsilon^{ijk}\Gamma_{jka}+\gamma K^i_a$ known as the Ashtekar-Barbero connection \cite{Ashtekar:1986yd,Barbero:1994ap}. Here, $\Gamma_{jka}$ is the spin connection, $K^i_a:=\Gamma^i_{\ 0a}$ is the dreibein-contracted extrinsic curvature, and $\gamma$ denotes the Immirzi parameter. By neglecting the possible coupling of spinors to gravity, the kinematics of the gravitational field can be defined by the Einstein-Hilbert action
\begin{align}\label{Einstein-Hilbert}
S^{(\text{gr}\@)}:=\frac{1}{\kappa}\!\int_{\!M}\!\!d^{4}x\sqrt{-g}R\,,
\end{align}
where $R$ is the Ricci scalar and the gravitational coupling constant reads $\kappa=16\pi G$. The momentum of  $A^i_a$ is given by the densitized dreibein $E_i^a:=\sqrt{q}e_i^a$. These fields are in the canonical relation
\begin{align}\label{Poisson}
\big\{A^i_a(t,{\bf x}), E_j^b(t,{\bf y})\big\}=\frac{\gamma\kappa}{2}\delta_a^b\,\delta_j^i\,\delta^{3}({\bf x}-{\bf y})
\end{align}
with respect to the ADM variables.

The total Hamiltonian $H^{(A)}_T=G^{(A)}+V^{(A)}+H^{(A)}$ corresponding to \eqref{Einstein-Hilbert} is composed of three constraints
\begin{align}
\begin{split}
G^{(\text{gr}\@)}:=&\;\frac{1}{\gamma\kappa}\!\int_{\Sigma_t}\!\!\!\!d^3x\,A^i_t
D_aE^a_i\,,
\\
V^{(\text{gr}\@)}
:=&\;\frac{1}{\gamma\kappa}\!\int_{\Sigma_t}\!\!\!\!d^3x\,N^a
F^i_{ab}E^b_i\,,
\\
H^{(\text{gr}\@)}
:=&\;\frac{1}{\kappa}\!\int_{\Sigma_t}\!\!\!\!d^3x\,N
\bigg(
\frac{1}{\sqrt{q}}\big(F^i_{ab}-(\gamma^2+1)\epsilon_{ilm}K^l_aK^m_b\big)\epsilon^{ijk}E_j^aE_k^b
\bigg).
\end{split}
\end{align}
The operator $D_a$ is the covariant derivative of the Ashtekar-Barbero connection and the curvature of this $\mathfrak{su}(2)$ field is specified by $F^i_{ab}:=\partial_aA^i_b-\partial_bA^i_a+\epsilon_{ijk}A_a^jA_b^k$.

As in the case of the matter vector field, the constraints that do not contain propagating degrees of freedom are solved classically. These are the Gauss constraint $G^{(A)}$ and the diffeomorphism one $V^{(A)}$. The scalar constraint is regularized and quantized on the $\mathfrak{su}(2)$-invariant and diffeomorphisms-invariant lattice. These procedures lead to the HCO for gravity. The fields $F^i_{ab}$ and $E_i^a$ are regularized in the way presented in \eqref{YMWilson} and \eqref{YMflux}, respectively. Due to the internal symmetry of the Ashtekar variables, the precise formulas read
\begin{align}
\label{GRWilson}
\text{tr}(\tau^ih_{l\circlearrowleft l'})
=&\,-\frac{1}{2}\varepsilon_l\varepsilon_{l'}\und{F}^i_{ab}\dot{l}^a\dot{l}'^b+\mathcal{O}(\varepsilon^3)
\intertext{and}
\label{GRflux}
f_i(\mathbf{S}):=&\int_{\mathbf{S}}\!n_aE^{a}_i\,,
\end{align}
respectively.

The object absent in \eqref{YMscalar} but present in $H^{(\text{gr}\@)}$ is the extrinsic curvature $K^i_a$. However, in the cosmological framework discussed in this article, the spin connection $\Gamma_{jka}$ contribution to the constant field $A^i_a$ vanishes and this latter field becomes proportional to $K^i_a$. Therefore any separated regularization of the extrinsic curvature does have to be introduced. Finally, the lattice smearing of the Ashtekar-Barbero connection is defined in analogy to \eqref{GRWilson},
\begin{align}\label{GRholonomy}
\text{tr}\big(\tau^ih_{l}\big)=-\frac{1}{2}\varepsilon_l\und{A}_a^i\dot{l}^a+\mathcal{O}(\varepsilon^2)\,.
\end{align}
It is worth noting that expressions \eqref{GRWilson} and \eqref{GRholonomy} are not expended up to the same order. We neglect this problem in this article, although it is an essential issue concerning the general investigation of the lattice regularization procedure in LQG, \textit{cf.} \cite{Bilski:2020poi,Bilski:2020xfq}.

\subsection{Methods of scalar fields coupling to LQG}\label{II.4}

\noindent
The simplest classical representative of the bosonic matter content in cosmology is the real massless scalar field $\varphi$ without internal degrees of freedom. To formulate a diffeomorphism-invariant representation of $\varphi$, one needs to rely on a different strategy than for vector fields. This issue is related to different geometrical properties of the aforementioned objects. The scalar field and its momentum $\pi$ are not a one-form density and a vector density, respectively, but a scalar and a pseudoscalar density, respectively. Therefore, their smearing along a link and through a surface would not be correct. The point-solid duality appears to be the right pair of objects that allows one to describe the degrees of freedom of $\varphi$ and $\pi$ on a lattice.

The massless Klein-Gordon scalar field is defined by the action
\begin{align}\label{singlet_action}
S^{(\varphi)}:=\frac{1}{2\mathsf{g}_{\varphi}^2}\!\int_{\!M}\!\!\!d^{4}x\sqrt{-g}\,g^{\mu\nu}\partial_{\mu}\varphi\,\partial_{\nu}\varphi\,,
\end{align}
where  $\mathsf{g}_{\varphi}^2$ is the coupling constant. The Legendre transform results in the same structure of the total Hamiltonian $H^{(\varphi)}_T=V^{(\varphi)}+H^{(\varphi)}$ as in the case of the vector matter field. The diffeomorphism and Hamiltonian analogs of the constraints in \eqref{YMvector} and \eqref{YMscalar} are
\begin{align}\label{scalar_vector}
V^{(\varphi)}:=&\;\int_{\Sigma_t}\!\!\!\!d^3x\,N^a\partial_a\varphi\,\pi
\intertext{and}
\label{scalar_scalar}
H^{(\varphi)}:=&\;\frac{1}{2}\!\int_{\Sigma_t}\!\!\!\!d^3x\,N
\bigg(\frac{\mathsf{g}_{\varphi}^2}{\sqrt{q}}\pi^2+\frac{\sqrt{q}}{\mathsf{g}_{\varphi}^2}q^{ab}\partial_a\varphi\,\partial_b\varphi\bigg),
\end{align}
respectively. The explicit form of the momentum canonically conjugated to $\varphi$ is $\pi=\frac{\sqrt{q}}{\mathsf{g}_{\varphi}^2}e_0^{\mu}\partial_{\mu}\varphi$. It is worth noting that the aforementioned quantities can be easily extended to the self-interacting field formalism. In this case, the potential can be given by a polynomial of $\varphi$. This potential trivially couples to gravity only by multiplication with $\sqrt{q}$, hence it does not bring any significant contribution to the analysis in this article.

The simplest point-solid symmetry reflecting the lattice representation of the Klein-Gordon field is the following. The holonomy-like representation \cite{Thiemann:1997rt,Thiemann:1997rq} located at a node $v$ (an intersection of links) is
\begin{align}
\label{point_holonomy}
\Phi_v:=\exp\!\big(\i\varepsilon_{\@v}\varphi(x)\big)\,.
\end{align}
The solid-related momentum representation is
\begin{align}
\label{point_momentum}
\Pi(R_v):=\int_{\!R_v}\!\!\!d^3x\.\delta_{v,x}\.\pi(x)\,,
\end{align}
where the integration was done all over the region $R_v$, centered at $v$. The last quantity in \eqref{point_momentum} is assumed to be \textit{a priori} smeared, reading $\pi(x):=\sum_{y\in R}\!\delta^3\@(x\@-\@y)\.\Pi(y)$.

The preceding pair of definitions is related to nodes. Their trivial distribution all over the lattice leads to a simple construction of the related Fock space. In this case one usually considers the polymer representation \cite{Ashtekar:2002vh,Kaminski:2005nc,Kaminski:2006ta,Hossain:2010wy} with nodes-located states having a similar form to the definition in \eqref{point_holonomy}.

\subsection{Models contradictive with CQGR}\label{II.5}

\noindent
At the end of this section three popular quantum cosmological models that are indirectly related to LQG are going to be discussed. Each of these models is associated with a different approach to regularize and quantize matter. By following the review of these theories \cite{Wilson-Ewing:2016yan}, three quantization procedures can be recognized: the effective constraints, dress metric, and the separate universe quantization approach. By concerning the methodology introduced in Sec.~\ref{I}, one can verify if these models meet the quantum general postulate of relativity criterion. In this way, one can check if any of these approaches could be considered as a simplification of CQGR, hence as a phenomenology-valued cosmological model.

The effective constraints method \cite{Bojowald:2012xy,Barrau:2013ula,Barrau:2014maa} does not define any QFT for matter. Instead, it introduces unspecified perturbations around the classical cosmological matter density derived from LQC \cite{Ashtekar:2003hd,Bojowald:2008zzb,Ashtekar:2011ni}. The structure of these perturbations is then restricted by a closeness of the constraint algebra. Therefore, this effective model is not \textit{a priori} contradictive with CQGR, unless the formulation of LQC is not a cosmological limit of CQGR. In the latter case, one can always repeat the procedures of the effective constraints method around a different cosmological framework obtained from LQG. However, the results of this model, by definition, do not provide any insight into the structure of the matter sector. This approach formulates only an effective description of the cosmological data, but it does not describe the mechanism that explains the origin of this data. Hence, the effective constraints method may have physical applications but not as a phenomenology predicting technique.

The dressed metric approach is based on the idea proposed in \cite{Ashtekar:2009mb}. It was applied both to the scalar \cite{Agullo:2012fc} and vector \cite{Lewandowski:2017cvz} fields, which were described by the method of QFT in curved spacetime. By defining the Fock space for matter fields and by choosing the expectation value of the HCO in LQC as a background, one directly violates SE in the construction of the theory. As a consequence, the approximation of this model omits the corrections that otherwise would be present as a result of the quantization of the gravitational degrees of freedom in the HCO of the matter sector. These corrections would be of the same order of significance as the cosmological corrections from the free sector of gravity and the QFT perturbations of matter --- see also Sec.~\ref{IV.1}. This argument demonstrates that the BI violation indicates the related incompleteness of the results. A particular form of the corrections based on the SE formulation of LQC and the BI method of the derivation of its semiclassical results is used to define the background on which the Fock space for the matter sector is constructed. Then, by definition, these corrections will be absent in the semiclassical limit of the matter sector. Therefore, the dressed metric approach is an inconsistently formulated toy model and it cannot be applied as a physical tool. This model can be used only to study particular theoretical mechanisms. It is worth noting that a specific variant of this approach, called the hybrid quantization \cite{FernandezMendez:2012vi,Gomar:2014faa}, additionally assumes left and right multiplication of the total HCO by the quantized equivalent of the $q^{-1/4}$ quantity. This affects the gravitational sector, which, by construction of the mentioned multiplication, cannot be thought as a limit of SE LQG. In the case of a possibility to compare any results of this model with data, the same predicable inapplicability arguments as in the case of the dressed metric approach hold.

Finally, the separate universe quantization \cite{Wilson-Ewing:2015sfx} is the the long-wavelength gravitational modes quantization on the LQC background. This method uses the long-wavelength approximation to construct a loop quantization both for the background and perturbations. It could be an improvement of the dressed metric approach for particular applications. This model does not assume a separate quantization for the background variables (in the LQG-like method) and the perturbative degrees of freedom (in the Fock space method) like the previously discussed approach. However, the separate universe quantization generates a different problem by neglecting the specific structure of quantum matter fields with their corresponding corrections. It is difficult to imagine a generalization of this effective approach to all the matter fields in the Standard Model of particle physics. Moreover, from the perspective of a simple effective cosmological theory, the separate universe quantization neglects the gravitational corrections to the matter sector. Hence, by construction, this model cannot be expected to be the cosmological limit of a fundamental theory of CQGR, which would be constructed as LQG with an analogous quantization of matter fields.

Concluding, all the aforementioned models are not fundamental and cannot provide any certain insight into real cosmological processes. They are not compatible with the SE and BI canonical procedures of QFT (including the theory of gravity) on a lattice. However, they can be used to study particular theoretical or mathematical procedures. Moreover, the first one, the effective constraints method can be used to describe all the cosmological data statistically. The possible physical application of the second and the third model is more limited. However, they can still be used as the effective tools to describe cosmological data from the epochs in which their incomplete predictions are expected to be negligible. In general, none of these models is expected to give a deeper insight into the understanding of cosmology than the standard methods of QFT on curved spacetime. Let it be emphasized that this statement is based on the assumption that the hypothetical fundamental quantum theory of gravity and matter has SE and BI. More detailed critical reviews concerning also other problems of the mentioned models can be found in \cite{Bojowald:2020xlw,Bojowald:2020wuc,Bojowald:2020unm}.

In the next section a toy model satisfying the SE condition will be constructed. Then its BI will be tested to set a direction toward future attempts of a fundamental model construction.


\section{Kinematics}\label{III}

\subsection{Cosmological models}\label{III.1}

\noindent
To discuss general covariance regarding the semiclassical limit of what could be a SE cosmological limit of LQG methods-based fundamental theory, one needs to consider consistent regularization and quantization procedures. The cosmological phase space reduction of the lattice-regularized gravity formulated in the Ashtekar variables is described in \cite{Bilski:2019tji}. Here, the cosmological phase space reduction is defined as the SU$(2)$ breakdown of the internal space invariance into the U$(1)$ case and the breakdown of the spatial diffeomorphisms into the ones that satisfy the Bianchi I symmetry. This result of the reduction is identical if it is done before the regularization and if after the reduction, the corresponding lattice structure is adjusted to the symmetry of the reduced Ashtekar variables \cite{Bilski:2019tji}. In both cases of the phase space reduction, applied either to the holonomy-flux description or the Ashtekar variables formulation, the resulting (classical) lattice-regularized Hamiltonian constraint is equivalent with the one assumed in the Bianchi I extension of LQC in \cite{Ashtekar:2009vc}. The reduction can be also applied to the expectation values of the operators on the states providing the classical limit of the system. Naturally, the structure of the HCO is again the same \cite{Bilski:2019tji}. In this latter case, however, the states are already given by the formalism of LQG (these are the symmetry-reduced spin network states  \cite{Thiemann:1996aw,Thiemann:1996av,Thiemann:1997rv}) and are different than the ones assumed in the extended LQC \cite{Ashtekar:2009vc,Ashtekar:2011ni}.

By assuming the expectation value of the HCO for the cosmological reduction of LQG and by keeping all the quantum corrections up to the quadratic order in the regularization parameter $\varepsilon$, one obtains
\begin{align}
\label{sinsin}
\big<\hat{H}^{(\text{gr}\@)}\big>=
-\frac{2}{\gamma^2\kappa}\sum_vN_v\sqrt{\bar{E}^a_1\delta_a^1\bar{E}^b_2\delta_b^2\bar{E}^c_3\delta_c^3}
\,\frac{1}{|\bar{E}^d_i\delta_d^i|}
\prod_{k\neq i}\frac{\sin\@\big(\varepsilon_{(\@k\@)}\.\bar{A}_e^k\delta^e_k\@\big)}{\varepsilon_{(\@k\@)}\!}
\Bigg[1+\mathcal{O}\Bigg(\frac{1}{\!\big(\bar{j}^{(i)}\@\big)^{\!2\@}}\!\Bigg)\Bigg].
\end{align}
Here, $\bar{A}_a^{(i)}\delta^a_i$ is the Ashtekar connection's diagonal sector that is obtained by a simultaneous fixing of the internal and diffeomorphism symmetries. Analogously, $\bar{E}^a_i$ denotes the diagonal densitized dreibein, reintroduced by the correspondence principle, \textit{i.e.} by replacing the eigenvalue of the $\hat{E}^a_i$ operator with its classical equivalent. It is worth mentioning that the same result was postulated by the incorrectly derived \cite{Bilski:2019tji} `partial quantum reduction procedure' known as quantum reduced loop gravity (QRLG) \cite{Alesci:2013xd,Alesci:2014uha}. Moreover, the expectation value of HCO in cosmological coherent quantum gravity (CCQG) \cite{Dapor:2017gdk} derived from LQG by assuming a particular selection of states is the same up to the terms of order $\varepsilon^2$. Moreover, a similar expression with the same structure up to the order $\varepsilon^2$ appears when the Lorentzian term (see the subsequent analysis) is taken into account \cite{Yang:2009fp}. Finally, the second term in the quadratic bracket in \eqref{sinsin} is not written explicitly because it differs in the mentioned models by a numerical factor. This term is known as the inverse volume corrections. It was first noticed in the isotropic model of LQC in \cite{Bojowald:2001vw}. As demonstrated in \cite{Grain:2009cj}, the related corrections have a significant contribution to the dynamics of the primordial universe. Their structure is going to be the essential issue studied in this article.

By neglecting  the differences in the order-$\varepsilon^3$ corrections coming from the expansion of the trigonometric functional in the gravitational sector of the scalar constraint\footnote{The form of the trigonometric functional slightly varies from one model to another. However, this form remains always expandable into a power series of connections.}, the structure of the expectation value of the HCO in the aforementioned models, including the inverse volume corrections, remains the same up to a constant factor\footnote{This statement is true in general, as long as the volume operator is an eigenoperator of the states defined on a cuboidal lattice, \textit{cf.} \cite{Flori:2008nw}.}. These latter corrections appear as a result of the action of the operator
\begin{equation}
\label{inverse}
\hat{h}_a^{-1}\big[\hat{\mathbf{V}},\hat{h}_a\big]\,.
\end{equation}
Analogous corrections to the gravitational degrees of freedom are present also in the matter sector. These corrections are sourced from a more general expression
\begin{equation}
\label{inverse_general}
\hat{h}_a^{-1}\big[\hat{\mathbf{V}}^n,\hat{h}_a\big]\,,
\end{equation}
where $n$ is a positive rational number. It is worth being emphasized that here the volume operator acts on a state that is initially modified by the gravitational holonomy operator $\hat{h}_a$. The latter operator acts by multiplication. Then the difference in the states on which the volume operator acts, results in the inverse volume corrections.

To identify each occurrence of the terms given in \eqref{inverse_general}, as well as each value of the $n$ parameter in the HCO of the entire system, it is enough to investigate all the classical contributions to the lattice-smeared scalar constraint. The explicit expression of this object for the torsionless gravity is given by
\begin{align}
\label{H_grav}
H^{(\text{gr}\@)}
=\int_{\Sigma_t}\!\!\!\!d^3x\.N\@(x)
\Big(
\mathcal{H}^{(\text{gr}\@)}_{_\text{Eucl}\@}\@(x)+\mathcal{H}^{(\text{gr}\@)}_{_\text{Lor}}\@(x)
\Big)\,,
\end{align}
where
\begin{align}
\label{H_grav_Eucl}
\mathcal{H}^{(\text{gr}\@)}_{_\text{Eucl}\@}\@(x)
:=&\;
\frac{2^2}{\gamma\kappa^2}\lim_{\varepsilon\to0}
\epsilon^{abc}
\,\text{tr}\bigg(
\frac{1}{\varepsilon^2}\Big(h_{ab}^{\mathstrut}\@(x)-h_{ab}^{-1}\@(x)\Big)
\frac{1}{\varepsilon}h_c^{-1}\@(x)\Big\{\mathbf{V}\@(x),h_c^{\mathstrut}\@(x)\Big\}
\bigg)
\intertext{and}
\label{H_grav_Lor}
\mathcal{H}^{(\text{gr}\@)}_{_\text{Lor}\@}\@(x)
:=&\;
-\frac{2^5(\gamma^2\@+\@1)}{\gamma^3\kappa^4}\lim_{\varepsilon\to0}
\epsilon^{abc}
\,\text{tr}\bigg(
\frac{1}{\varepsilon}h_a^{-1}\@(x)\Big\{\mathbf{K}\@(x),h_a^{\mathstrut}\@(x)\Big\}
\frac{1}{\varepsilon}h_b^{-1}\@(x)\Big\{\mathbf{K}\@(x),h_b^{\mathstrut}\@(x)\Big\}
\frac{1}{\varepsilon}h_c^{-1}\@(x)\Big\{\mathbf{V}\@(x),h_c^{\mathstrut}\@(x)\Big\}
\bigg).
\end{align}
The terms in the form of the expression in \eqref{inverse} are easily recognizable.

The structure of the lattice corrections in the matter sector is also directly readable from the regularized form of the Hamiltonian constraint. In the case of the vector field, it is given by
\begin{align}
\label{H_vect}
H^{(\underline{A})}
=\int_{\Sigma_t}\!\!\!\!d^3x\.N\@(x)
\Big(
\mathcal{H}^{(\underline{A})}_{_\text{elec}\mathstrut\!}\@(x)+\mathcal{H}^{(\underline{A})}_{_\text{magn}\mathstrut\!}\@(x)
\Big)\,,
\end{align}
where
\begin{align}
\label{H_vect_elec}
\begin{split}
\mathcal{H}^{(\underline{A})}_{_\text{elec}\mathstrut\!}\@(x)
=&\;
\frac{2^7\mathsf{g}_{\!\und{A}}^2}{(\gamma\kappa)^2}\lim_{\varepsilon\to0}
\underline{E}^a\@(x)
\,\text{tr}\bigg(
\tau^i\frac{1}{\varepsilon}h_a^{-1}\@(x)\Big\{\mathbf{V}^{\frac{1}{2}}\@(x),h_a^{\mathstrut}\@(x)\Big\}
\bigg)
\\
&\,\times
\!\int\!\!d^3y\.\delta^3\@(x\@-\@y)\.
\underline{E}^b\@(y)
\,\text{tr}\bigg(
\tau^i\frac{1}{\varepsilon}h_b^{-1}\@(y)\Big\{\mathbf{V}^{\frac{1}{2}}\@(y),h_b^{\mathstrut}\@(y)\Big\}
\bigg)
\end{split}
\intertext{and}
\label{H_vect_magn}
\begin{split}
\mathcal{H}^{(\underline{A})}_{_\text{magn}\mathstrut\!}\@(x)
=&\;
\frac{2^7\mathsf{g}_{\!\und{A}}^2}{(\gamma\kappa)^2}\lim_{\varepsilon\to0}
\underline{B}^a\@(x)
\,\text{tr}\bigg(
\tau^i\frac{1}{\varepsilon}h_a^{-1}\@(x)\Big\{\mathbf{V}^{\frac{1}{2}}\@(x),h_a^{\mathstrut}\@(x)\Big\}
\bigg)
\\
&\,\times
\!\int\!\!d^3y\.\delta^3\@(x\@-\@y)\.
\underline{B}^b\@(y)
\,\text{tr}\bigg(
\tau^i\frac{1}{\varepsilon}h_b^{-1}\@(y)\Big\{\mathbf{V}^{\frac{1}{2}}\@(y),h_b^{\mathstrut}\@(y)\Big\}
\bigg).
\end{split}
\end{align}
Analogously, the regularized scalar field Hamiltonian is expressed by
\begin{align}
\label{H_scal}
H^{(\varphi)}
=\int_{\Sigma_t}\!\!\!\!d^3x\.N\@(x)
\Big(
\mathcal{H}^{(\varphi)}_{_\text{mom}\mathstrut\!}\@(x)+\mathcal{H}^{(\varphi)}_{_\text{der}\mathstrut\!}\@(x)+\mathcal{H}^{(\varphi)}_{_\text{pot}\mathstrut\!}\@(x)
\Big)\,,
\end{align}
where
\begin{align}
\label{H_scal_mom}
\begin{split}
\mathcal{H}^{(\varphi)}_{_\text{mom}\mathstrut\!}\@(x)
=&\;
\frac{2^{21}\mathsf{g}_{\varphi}^2}{3^2(\gamma\kappa)^6}\lim_{\varepsilon\to0}
\pi\@(x)
\.\epsilon_{ijk}\.\epsilon^{abc}\!\int\!\!d^3z\.\delta^3\@(x\@-\@z)
\,\text{tr}\bigg(
\tau^i\frac{1}{\varepsilon}h_a^{-1}\@(z)\Big\{\mathbf{V}^{\frac{1}{2}}\@(z),h_a^{\mathstrut}\@(z)\Big\}
\bigg)
\\
&\,\times
\,\text{tr}\bigg(
\tau^j\frac{1}{\varepsilon}h_b^{-1}\@(z)\Big\{\mathbf{V}^{\frac{1}{2}}\@(z),h_b^{\mathstrut}\@(z)\Big\}
\bigg)
\,\text{tr}\bigg(
\tau^k\frac{1}{\varepsilon}h_c^{-1}\@(z)\Big\{\mathbf{V}^{\frac{1}{2}}\@(z),h_c^{\mathstrut}\@(z)\Big\}
\bigg)
\\
&\,\times
\!\int\!\!d^3y\.\delta^3\@(x\@-\@y)\.\pi\@(y)
\.\epsilon_{lmn}\.\epsilon^{def}\!\int\!\!d^3z'\.\delta^3\@(y\@-\@z')
\,\text{tr}\bigg(
\tau^l\frac{1}{\varepsilon}h_d^{-1}\@(z')\Big\{\mathbf{V}^{\frac{1}{2}}\@(z'),h_d^{\mathstrut}\@(z')\Big\}
\bigg)
\\
&\,\times
\,\text{tr}\bigg(
\tau^m\frac{1}{\varepsilon}h_e^{-1}\@(z')\Big\{\mathbf{V}^{\frac{1}{2}}\@(z'),h_e^{\mathstrut}\@(z')\Big\}
\bigg)
\,\text{tr}\bigg(
\tau^n\frac{1}{\varepsilon}h_f^{-1}\@(z')\Big\{\mathbf{V}^{\frac{1}{2}}\@(z'),h_f^{\mathstrut}\@(z')\Big\}
\bigg),
\end{split}
\\
\label{H_scal_der}
\begin{split}
\mathcal{H}^{(\varphi)}_{_\text{der}\mathstrut\!}\@(x)
=&\;\frac{2^{17}}{3^4(\gamma\kappa)^4\.\mathsf{g}_{\varphi}^2}\lim_{\varepsilon\to0}
\partial_a\varphi(x)
\.\epsilon_{ijk}\.\epsilon^{abc}
\,\text{tr}\bigg(
\tau^j\frac{1}{\varepsilon}h_b^{-1}\@(x)\Big\{\mathbf{V}^{\frac{3}{4}}\@(x),h_b^{\mathstrut}\@(x)\Big\}
\bigg)
\,\text{tr}\bigg(
\tau^k\frac{1}{\varepsilon}h_c^{-1}\@(x)\Big\{\mathbf{V}^{\frac{3}{4}}\@(x),h_c^{\mathstrut}\@(x)\Big\}
\bigg)
\\
&\,\times
\!\int\!\!d^3y\.\delta^3\@(x\@-\@y)\.\partial_d\varphi(y)
\.\epsilon_{ilm}\.\epsilon^{def}
\,\text{tr}\bigg(
\tau^l\frac{1}{\varepsilon}h_e^{-1}\@(y)\Big\{\mathbf{V}^{\frac{3}{4}}\@(y),h_e^{\mathstrut}\@(y)\Big\}
\bigg)
\,\text{tr}\bigg(
\tau^m\frac{1}{\varepsilon}h_f^{-1}\@(y)\Big\{\mathbf{V}^{\frac{3}{4}}\@(y),h_f^{\mathstrut}\@(y)\Big\}
\bigg)
\end{split}
\intertext{and}
\label{H_scal_pot}
\mathcal{H}^{(\varphi)}_{_\text{pot}\mathstrut\!}\@(x)
=&\;\frac{1}{2\.\mathsf{g}_{\varphi}^2}
\sqrt{\@q(x)}\.\mathcal{V}[\varphi(x)]
\approx\frac{1}{2\.\mathsf{g}_{\varphi}^2}\lim_{\varepsilon\to0}
\mathcal{V}[\varphi(x)]\.\frac{1}{\varepsilon^3\!}\mathbf{V}\@(x,\varepsilon)\,.
\end{align}
Here, for clearness, the potential term was introduced to demonstrate the ambiguity in the choice of its form $\mathcal{V}[\varphi(x)]$ about the presence of the related gravitational corrections. These do not appear because, after the quantization, the volume operator does not act on holonomy-modified states.

\subsection{States space}\label{III.2}

\noindent
To discuss the form of the semiclassical quantum-geometrical corrections in the cosmological simplification of CQGR one needs to specify the states space. It is worth repeating the fact already recalled in the previous subsection. The only terms in the scalar constraint contributing to the next-to-the-leading-order inverse volume corrections have the same structure independently of the selected cosmological model that does not break SE. These terms depend entirely on the postulated classical action of all the contributing fields and on the power of volume in the following approximate identity \cite{Thiemann:1996aw,Thiemann:2007zz,Thiemann:1997rt} applied to the regularization procedure,
\begin{align}
\label{E_trick}
\big\{A^i_a,\mathbf{V}^{n}\big\}=\frac{n\gamma\kappa}{4}\frac{1}{E_i^a}\big(\sqrt{|E|}\big)^{\!n}+\mathcal{O}(\varepsilon)\,.
\end{align}
It is worth noting that neglecting the last lattice correction term is as precise as neglecting the analogous correction in \eqref{GRholonomy}. Moreover, in the $n=1$ case this term vanishes identically and the additional constraint $\varepsilon n\ll 1$ is required. Furthermore, in the limit $\varepsilon\to0$, this correction vanishes and the whole lattice-regularized system takes an $\varepsilon$-independent finite form.

The fact that the structure of next-to-the-leading-order inverse volume corrections does not depend on the selected cosmological formulation is a result of the proper phase space reduction of the hypothetical fundamental theory. The reduction of variables into the Bianchi I symmetry have to entail the reduction of the lattice structure into the cuboidal form \cite{Bilski:2019tji}. The volume operator or its power $\mathbf{V}^{n}$, expressed as a functional of the diagonal densitized dreibein fluxes, is an eigenoperator of the states defined on a cuboidal lattice \cite{Flori:2008nw}. The modifications of the states by the holonomy contribution to formula \eqref{inverse_general} generate the inverse volume corrections along directions of these holonomies. Therefore, to investigate the semiclassical structure of the generated inverse volume corrections, it is enough to select the simplest states that reveal these corrections and derive the semiclassical limit. This last step can be easily done by defining the coherent states as the states that restore the volume from the eigenvalue of the related operator $\hat{\mathbf{V}}$ by the correspondence procedure.

One can consider the system of minimally coupled bosonic matter and gravity with the Hilbert space
\begin{align}
\label{total_space}
\mathscr{H}_{kin}\!:=\mathscr{H}_{kin}^{(\text{gr}\@)}\!\otimes\mathscr{H}_{kin}^{(\underline{A})}\!\otimes\mathscr{H}_{kin}^{(\varphi)}\,.
\end{align}
The vector matter field sector is labeled by $\mathscr{H}_{kin}^{(\underline{A})}$ and it is defined analogously to the one for the SU$(2)$-invariant gravitational field in LQG labeled by $\mathscr{H}_{kin}^{(\text{gr}\@)}$, \textit{cf.} \cite{Thiemann:1996aw,Thiemann:2007zz}. In both cases one assumes the space of cylindrical functions of the gauge connections holonomies. Also, in both cases, the basis states are the invariant spin network states: $|\Gamma;\underline{j}_l\rangle$ for the Abelian vector field and $|\Gamma;j_l,i_v\rangle$ for the SU$(2)$-invariant gravitational field, respectively. They are labeled by quantum numbers (spins) $\underline{j}_l$ and $j_l$, respectively. These numbers determine the notion of the gauge groups irreducible representations at each link $l$. To preserve the gauge invariance in the non-Abelian gravitational case of LQG, the corresponding intertwiners $i_v$ are attached at each node $v$. The reduced phase space approach allows one to fix the internal space to the Abelian U$(1)$ case \cite{Bilski:2019tji}. Consequently, one can drop the trivial intertwiners from states; any Hilbert space is, by definition, specified up to a number. Concluding, the simplest states for the matter and gravitational vector fields are defined along the cuboidal lattice links and are denoted by $|\Gamma;\underline{j}_l\rangle\in\mathscr{H}_{kin}^{(\underline{A})}$ and $|\Gamma;j_l\rangle\in\mathscr{H}_{kin}^{(\text{gr}\@)}$.

The nodes-related states are qualitatively different. The Hilbert space describing the scalar field point holonomy representation is defined as
\begin{equation}
\mathscr{H}_{kin}^{(\varphi)}:=\overline{\big\{a_1U_{\pi_1}+...+a_nU_{\pi_n}\!:\ a_i\in\mathds{C},\, n\in\mathds{N},\, \pi_i\in\mathds{R}\big\}}\,,
\end{equation}
where the wave function reads
\begin{equation}
U_\pi(\varphi):=\langle\varphi|U_{\{v_1,..,v_n\},\{\pi_{v_1},\pi_{v_n}\}}\rangle:=e^{\i\sum_{v\in\Sigma}\pi_v\varphi_v}\,.
\end{equation}
This definition explicitly preserves the rotational symmetry of the scalar field at each point. The whole collection of nodes forms a trivial polymerlike structure. Moreover, by construction, this structure is diffeomophically independent of any spacetime geometry. The modes of the scalar field do not oscillate in space but are statically located at points and the distance between these points is only trivially coupled with gravity --- by multiplication. This isotropic structure distributed over the lattice does not reflect any possible internal quantum relation between the gravitational and matter degrees of freedom.

By considering the single-point state $\langle{\varphi}|v;U_{\pi}\rangle:=e^{\i\pi_v\varphi_v}$ located at $v$, the action of the canonical operators is trivially defined in the exponential form. The point holonomy shifts the state as follows:
\begin{equation}\label{canonical_field}
e^{\i\pi'_{v'}\hat{\varphi}_{v'}}|v;U_{\pi}\rangle:=e^{\i\pi'_{v'}\varphi_{v'}}|v;U_{\pi}\rangle=|v\cup{v'};U_{\pi+\pi'}\rangle\,.
\end{equation}
Analogous action for the momentum operator corresponding to the $\varepsilon$-smeared momentum $\pi$ in the region around the $v$ node is given by the eigenequation
\begin{equation}\label{canonical_momentum}
\hat{\Pi}({v'})|v;U_{\pi}\rangle:=-\i\hbar\frac{\partial}{\partial\varphi({v'})}|v;U_{\pi}\rangle=\hbar\pi_{v'}\delta_{v,{v'}}|v;U_{\pi}\rangle\,.
\end{equation}
The scalar product definition is simply adjusted to the trivial form of the canonical operators, reading
\begin{equation}
\langle v;U_{\pi}|{v'};U_{\pi'}\rangle=\delta_{v,{v'}}\delta_{\pi_v^{},\pi'_{v'}}\,.
\end{equation}
More details concerning these polymer states for the point holonomy representation are given in \cite{Bilski:2015dra,Ashtekar:2002vh,Kaminski:2005nc,Kaminski:2006ta,Hossain:2010wy}.

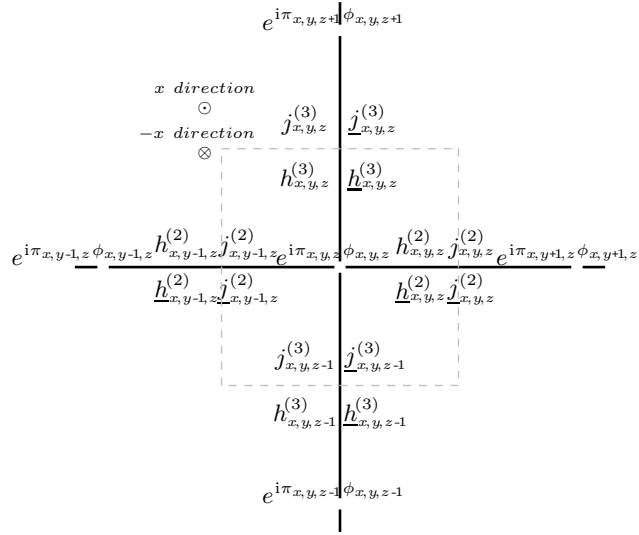
\begin{figure}[h]
\begin{center}
\begin{tikzpicture}[scale=0.75]
\node at (-2.7,0.4) {$h_{^{x\!,y\texttt{-\!}1\!,z}}^{\!(2)}$};
\node at (-1.6,0.4) {$j_{^{x\!,y\texttt{-\!}1\!,z}}^{(2)}$};
\node at (1.45,0.4) {$h_{^{x\!,y\!,z}}^{\!(2)}$};
\node at (2.35,0.4) {$j_{^{x\!,y\!,z}}^{(2)}$};
\node at (-0.6,1.6) {$h_{^{x\!,y\!,z}}^{\!(3)}$};
\node at (-0.6,2.6) {$j_{^{x\!,y\!,z}}^{(3)}$};
\node at (-0.6,-2.6) {$h_{^{x\!,y\!,z\texttt{-\!}1}}^{\!(3)}$};
\node at (-0.6,-1.6) {$j_{^{x\!,y\!,z\texttt{-\!}1}}^{(3)}$};
%
%
\node at (-2.7,-0.4) {$\underline{h}_{^{x\!,y\texttt{-\!}1\!,z}}^{\!(2)}$};
\node at (-1.6,-0.4) {$\underline{j}_{^{x\!,y\texttt{-\!}1\!,z}}^{(2)}$};
\node at (1.45,-0.4) {$\underline{h}_{^{x\!,y\!,z}}^{\!(2)}$};
\node at (2.35,-0.4) {$\underline{j}_{^{x\!,y\!,z}}^{(2)}$};
\node at (0.6,1.6) {$\underline{h}_{^{x\!,y\!,z}}^{\!(3)}$};
\node at (0.6,2.6) {$\underline{j}_{^{x\!,y\!,z}}^{(3)}$};
\node at (0.65,-2.6) {$\underline{h}_{^{x\!,y\!,z\texttt{-\!}1}}^{\!(3)}$};
\node at (0.65,-1.6) {$\underline{j}_{^{x\!,y\!,z\texttt{-\!}1}}^{(3)}$};
%
%
\draw[c] (-4.1,0) -- (-0.1,0);
\draw[c] (0.1,0) -- (4.1,0);
\draw[c] (0,0.1) -- (0,4.1);
\draw[c] (0,-0.1) -- (0,-4.1);
%
%
\draw[c] (-4.7,0) -- (-4.3,0);
\draw[c] (4.3,0) -- (4.7,0);
\draw[c] (0,4.3) -- (0,4.7);
\draw[c] (0,-4.7) -- (0,-4.3);
%
%
\node at (-0.1,4.4) {$e^{\i\pi_{x\!,y\!,z\texttt{+\!}1}\phi_{x\!,y\!,z\texttt{+\!}1}}$};
\node at (-4.4,0.22) {$\!\!\!\!e^{\i\pi_{x\!,y\texttt{-\!}1\!,z}\phi_{x\!,y\texttt{-\!}1\!,z}}$};
\node at (-0.1,0.22) {$e^{\i\pi_{x\!,y\!,z}\phi_{x\!,y\!,z}}$};
\node at (4.05,0.22) {$e^{\i\pi_{x\!,y\texttt{+\!}1\!,z}\phi_{x\!,y\texttt{+\!}1\!,z}}$};
\node at (-0.1,-4.0) {$e^{\i\pi_{x\!,y\!,z\texttt{-\!}1}\phi_{x\!,y\!,z\texttt{-\!}1}}$};
%
%
\draw[d] (-2.1,-2.1) -- (-2.1,2.1);
\draw[d] (-2.1,2.1) -- (2.1,2.1);
\draw[d] (2.1,2.1) -- (2.1,-2.1);
\draw[d] (2.1,-2.1) -- (-2.1,-2.1);
%
%
\node at (-2.4,3) {$\atopp{x\ direction}{\odot}$};
\node at (-2.4,2.2) {$\atopp{-x\ direction\ \ }{\otimes}$};
\end{tikzpicture}
\vspace{-5pt}
\caption{Normalized basic state of bosonic fields for cubic lattice}
\label{cellNORM}
\end{center}
\end{figure}
Concluding, the basis states are defined by
\begin{align}
\label{Fock}
\mathscr{H}_{kin}\ni|\Gamma;j_l,\underline{j}_l,U_{\pi}\rangle:=|\Gamma;j_l\rangle\otimes|\Gamma;\underline{j}_l\rangle\otimes|\Gamma;U_{\pi}\rangle\,.
\end{align}
By considering a single hexavalent node state $c_v\in\Gamma$, one can express the related Hilbert space structure in the graphical form, see FIG.~\ref{cellNORM}. The dashed frame specifies the normalization that allows to tessellate the reduced space with the embedded graph structure. This tessellation results in the set of the cuboidal cylindrical functions. Here, $j_{p,q,r}^{(i)}$ and $\underline{j}^{(i)}_{p,q,r}$ are the spin numbers associated to the links $l_{p,q,r}^{(i)}$. The scalar field state is represented by the point holonomy $e^{\i\pi_{p,q,r}\varphi_{p,q,r}}$ at the node $v_{p,q,r}\!\in\!\Gamma$, where $\pi_{p,q,r}$ is the real coefficient.

\begin{figure}[h]
\begin{center}
\begin{tikzpicture}[scale=0.75]
\node at (-2.7,0.4) {$h_{^{x\!,y\texttt{-\!}1\!,z}}^{\!(2)}$};
\node at (-1.6,0.4) {$j_{^{x\!,y\texttt{-\!}1\!,z}}^{(2)}$};
\node at (1.45,0.4) {$h_{^{x\!,y\!,z}}^{\!(2)}$};
\node at (2.35,0.4) {$j_{^{x\!,y\!,z}}^{(2)}$};
\node at (-0.6,1.6) {$h_{^{x\!,y\!,z}}^{\!(3)}$};
\node at (-0.6,2.6) {$j_{^{x\!,y\!,z}}^{(3)}$};
\node at (-0.6,-2.6) {$h_{^{x\!,y\!,z\texttt{-\!}1}}^{\!(3)}$};
\node at (-0.6,-1.6) {$j_{^{x\!,y\!,z\texttt{-\!}1}}^{(3)}$};
%
%
\node at (-2.7,-0.4) {$\underline{h}_{^{x\!,y\texttt{-\!}1\!,z}}^{\!(2)}$};
\node at (-1.6,-0.4) {$\underline{j}_{^{x\!,y\texttt{-\!}1\!,z}}^{(2)}$};
\node at (1.45,-0.4) {$\underline{h}_{^{x\!,y\!,z}}^{\!(2)}$};
\node at (2.35,-0.4) {$\underline{j}_{^{x\!,y\!,z}}^{(2)}$};
\node at (0.6,1.6) {$\underline{h}_{^{x\!,y\!,z}}^{\!(3)}$};
\node at (0.6,2.6) {$\underline{j}_{^{x\!,y\!,z}}^{(3)}$};
\node at (0.65,-2.6) {$\underline{h}_{^{x\!,y\!,z\texttt{-\!}1}}^{\!(3)}$};
\node at (0.65,-1.6) {$\underline{j}_{^{x\!,y\!,z\texttt{-\!}1}}^{(3)}$};
%
%
\draw[c] (-4.1,0) -- (-0.1,0);
\draw[c] (0.1,0) -- (4.1,0);
\draw[c] (0,0.1) -- (0,4.1);
\draw[c] (0,-0.1) -- (0,-4.1);
%
%
\draw[c] (-4.7,0) -- (-4.3,0);
\draw[c] (4.3,0) -- (4.7,0);
\draw[c] (0,4.3) -- (0,4.7);
\draw[c] (0,-4.7) -- (0,-4.3);
%
%
\node at (-0.1,4.4) {$e^{\i\pi_{x\!,y\!,z\texttt{+\!}1}\phi_{x\!,y\!,z\texttt{+\!}1}}$};
\node at (-4.4,0.22) {$\!\!\!\!e^{\i\pi_{x\!,y\texttt{-\!}1\!,z}\phi_{x\!,y\texttt{-\!}1\!,z}}$};
\node at (-0.1,0.22) {$e^{\i\pi_{x\!,y\!,z}\phi_{x\!,y\!,z}}$};
\node at (4.05,0.22) {$e^{\i\pi_{x\!,y\texttt{+\!}1\!,z}\phi_{x\!,y\texttt{+\!}1\!,z}}$};
\node at (-0.1,-4.0) {$e^{\i\pi_{x\!,y\!,z\texttt{-\!}1}\phi_{x\!,y\!,z\texttt{-\!}1}}$};
%
%
\draw[d] (-0.05,-0.05) -- (-0.05,4.15);
\draw[d] (-0.05,4.15) -- (4.15,4.15);
\draw[d] (4.15,4.15) -- (4.15,-0.05);
\draw[d] (4.15,-0.05) -- (-0.05,-0.05);
%
%
\node at (-2.4,3) {$\atopp{x\ direction}{\odot}$};
\node at (-2.4,2.2) {$\atopp{-x\ direction\ \ }{\otimes}$};
\end{tikzpicture}
\vspace{-5pt}
\caption{Basic state of bosonic fields for cubic lattice in LQC}
\label{spiderLQC}
\end{center}
\end{figure}
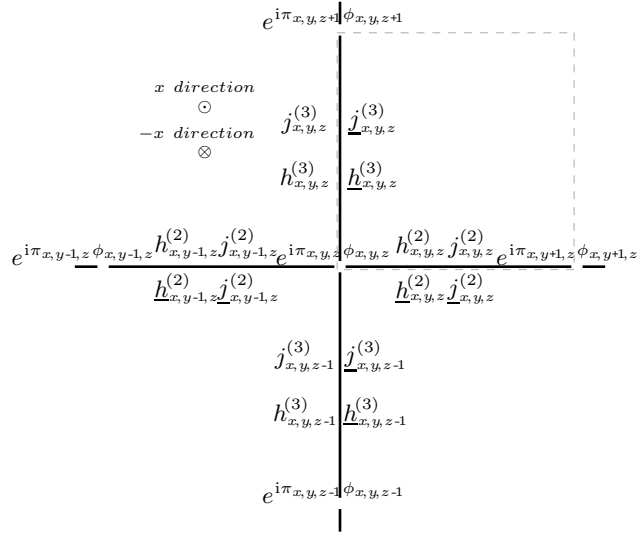
It is worth mentioning that the analogous structure in LQC is represented by a normalized hexavalent node state in FIG.~\ref{spiderLQC}. This latter structure also preserves the symmetry of the anisotropic Bianchi I model. However, it does not reflect the symmetry of the volume operator in LQG that acts at nodes. Moreover, the nodes-located distribution of the scalar field polymer structure has to coincide with the symmetry of any separable cellular form of the Fock space. This observation is based on the fact that the general form of the Fock space for LQG, known as the spin network, is not separable and has no \textit{a priori} defined centers of symmetry \cite{Thiemann:2007zz}. Conversely, the trivial states space for the scalar field restricts the modes oscillations to a point. To keep this symmetry at the quantum level, one cannot specify the elementary cell in the way proposed in FIG.~\ref{spiderLQC}. In this latter case, the scalar field contribution would need to be determined as the sum of the degrees of freedom at eight nodes, which would break the classical local rational symmetry of this field. Therefore, the specification of the elementary cell for the separable Fock space defined in \eqref{Fock} is uniquely restricted to the form in FIG.~\ref{cellNORM}.

Finally, only by staring at the structure of the state represented in FIG.~\ref{cellNORM}, one should recognize a methodological inconsistency in the construction of this multimatter coupling model --- the scalar field is lattice regularized in a qualitatively different manner than the other fields. On the one hand this inconsistency will be the source of the general covariance violation of the quantum corrections. On the other hand the consistency in the lattice regularization of the gravitational and the vector field will be preserved in the final results. This issue shows how to formulate the gravitational-matter coupling correctly in the future.

\subsection{Gravitational coherent states}\label{III.3}

\noindent
To discuss the semiclassical corrections precisely, one needs to define coherent states. For clearness of the analysis, the notation typical to LQC \cite{Ashtekar:2003hd,Ashtekar:2009vc,Ashtekar:2011ni} is going to be used. The reduced canonical variables \cite{Bilski:2019tji} are specified to 
\begin{align}
\label{homogeneous_A}
\tilde{A}^{i}_a(t):=&\;\frac{1}{l_0^{(i)}\!}\tilde{c}^{(i)\!}(t)\,{}^{0\!}e^{i}_a\,,
\\
\label{homogeneous_E}
\tilde{E}_{i}^a(t):=&\;\frac{l_0^{(i)}}{{\bf V}_{\!0}}\tilde{p}^{(i)\!}(t)\,\sqrt{^{0\!}q}\,{}^{0\!}e_{i}^a\,.
\end{align}
The matrices ${}^{0\!}e_{i}^a$ and ${}^{0\!}e^{i}_a$ represent constant orthonormal Cartesian frame and co-frame fields, respectively. The determinant of the fiducial metric ${}^{0\!}q_{ab}$ in \eqref{homogeneous_E} compensates the density weight of $\tilde{E}_{i}^a(t)$. The fiducial length $l_0^{(i)}$ and the corresponding volume ${\bf V}_{\!0}\!:=\!l_0^1l_0^2l_0^3$ of the fiducial cell are introduced to simplify the symplectic structure of the system. This leads to the following Poisson brackets,
\begin{align}
\label{Poisson_homogeneous}
\big\{\tilde{c}^{(i)}\@(t),\tilde{p}^{(j)}\@(t)\big\}=\frac{\kappa\gamma}{2}\delta^{\@(i)\@(j)}\,.
\end{align}

The semiclassical dynamics of the cosmologically reduced CQGR is specified by the Ehrenfest theorem\footnote{It is worth noting that the operator equation in \eqref{Ehrenfest}, known as the Ehrenfest theorem, was derived by Heisenberg \cite{Heisenberg:1929xj}.} and depends on the coherent states $|\ \rangle\in\mathscr{H}_{kin}$. The form of these states in this article is defined as the tensor product of the coherent states for different fields. The related Heisenberg equation reads,
\begin{align}
\label{Ehrenfest}
\frac{d}{dt}\big<\hat{O}\big>
-\bigg<\frac{\partial\hat{O}}{\partial t}\bigg>
=\frac{1}{\i\hbar}\big<[\hat{O},\hat{H}]\big>\,,
\end{align}
where the states factorize as follows:
\begin{align}
\label{coherent_tensor}
|\ \rangle=\widetilde{|\ \rangle\!}^{(\text{gr}\@)}\otimes|\ \rangle^{(\text{matt})}
=\widetilde{|\ \rangle\!}^{(\text{gr}\@)}\otimes\bigotimes_{\textbf{\straightphi}}|\ \rangle^{(\@\textbf{\straightphi}\@)}\,.
\end{align}
The symbol $\textbf{\straightphi}$ represents any matter field, and the term $\big<\frac{\partial\hat{O}}{\partial t}\big>$ was neglected by assuming only implicit time dependence of variables.

The normalized Bianchi I coherent states for the gravitational sector are defined as
\begin{align}
\label{coherent_gravity}
\widetilde{|\ \rangle\!}^{(\text{gr}\@)}:=
\sum_v\bigotimes_i^3\Big(
\big<\mathfrak{c}_{\@v\!}^{\@(\@i\@)\@\@}(\tilde{A})\big|\mathfrak{c}_{\@v\!}^{\@(\@i\@)\@\@}(\tilde{A})\big>
\Big)^{\@\@-\frac{1}{2}}
\big|\mathfrak{c}_{\@v\!}^{\@(\@i\@)\@\@}(\tilde{A})\big>\,.
\end{align}
The last factor is known as the shadow state \cite{Ashtekar:2002sn} with a $d$-width Gaussian distribution around the densitized dreibein operator eigenvalue. The form of this state reads
\begin{align}
\label{shadow_gravity}
\big|\mathfrak{c}_{\@v\!}^{\@(\@i\@)\@\@}(\tilde{A})\big>:=
\sum_{\mu^{(i)}_v}
\exp\!\Bigg[
{-\frac{1}{2d^2}\bigg(\frac{\mu^{(i)}_v\!}{2}\@-\frac{\tilde{p}^{(i)}\@}{\kbar}\bigg)^{\!\!2}}
\Bigg]
\exp\!\Bigg[
{-\i\bigg(\frac{\mu^{(i)}_v\!}{2}\@-\frac{\tilde{p}^{(i)}\@}{\kbar}\bigg)}\tilde{c}^{(i)}
\Bigg]\big|\mu^{(i)}_v\@\big>\,.
\end{align}
This formula is constructed on the link excitation states \cite{Ashtekar:2003hd,Ashtekar:2009vc} (the last factor above) that are given by the expression
\begin{align}
\label{shadow_basis}
\big|\mu^{(i)}_v\@\big>:=\exp\!\bigg[\i\frac{\mu^{(i)}_v\@}{2}\tilde{c}^{(i)}\bigg]\,,
\quad
\mu^{(i)}\in\mathds{Z}\,.
\end{align}
This definition is formulated in a direct analogy to the reduced form of the holonomy,
\begin{align}
\label{LQC_holonomy}
\tilde{h}_\nu^{(i)}(v):=\exp\!\bigg(\!\int_0^{\nu^{(i)}_vl_0^{(i)}}\!\!\!\!ds\,\tilde{A}_a^i\tau^i\dot{l}^a_{\nu}(s)\!\bigg)
=e^{\nu^{(i)}_v\tilde{c}^i\tau^i}\,,
\end{align}
\textit{cf.} \cite{Bilski:2019tji}.
Then, the actions of the lattice-regularized Bianchi I variables in \eqref{homogeneous_A} and \eqref{homogeneous_E} read
\begin{align}
\label{action_c}
\hat{\tilde{c}}^{(i)}\big|\mu^{(i)}_v\@\big>:=&\,-\frac{2}{\nu^{(i)}_v\@}\,\text{tr}\big(\tau^{(i)}\tilde{h}^{(i)}_\nu\big)\big|\mu^{(i)}_v\@\big>
=\frac{\i}{\nu^{(i)}_v\@}\Big(\big|\mu^{(i)}_v-\nu^{(i)}_v\big>-\big|\mu^{(i)}_v+\nu^{(i)}_v\big>\@\Big)
\intertext{and}
\label{action_p}
\hat{\tilde{p}}^{(i)}\big|\mu^{(i)}_v\@\big>:=&\,-\i\kbar\frac{\partial}{\partial\tilde{c}^{(i)}\@}\big|\mu^{(i)}_v\@\big>
=\frac{\mu^{(i)}_v\@}{2}\kbar\big|\mu^{(i)}_v\@\big>\,,
\end{align}
respectively.

In this article only the structure of the corrections, not their exact value, is going to be verified. This allows one to simplify the notation even more by replacing the reduced variables in \eqref{homogeneous_A} and \eqref{homogeneous_E} by
\begin{align}
\label{red_A}
\bar{A}^{i}_a(t):=&\;A^{(i)}_{(a)}\@(t)\,{}^{0\!}e^{i}_a=\frac{1}{\varepsilon\!}c^{(i)\!}(t)
\intertext{and}
\label{red_E}
\bar{E}_{i}^a(t):=&\;E_{(i)}^{(a)}\@(t)\,{}^{0\!}e_{i}^a=\frac{1}{\varepsilon^2\!}\.p^{(i)\!}(t)\,,
\end{align}
respectively, where $\varepsilon$ is the small regularization parameter. Here, the anisotropy and inhomogeneity of the regulators that depend on links lengths was neglected. This second simplification does not affect the structure of corrections. Moreover, in the properly reduced system, the final form of the Hamiltonian should be regulator-independent (this has been verified in the recent improved cosmological model in \cite{Bilski:2021_LCC}), hence this operation is not going to modify conclusions. The reduced holonomy becomes
\begin{align}
\label{red_holonomy}
h^{(i)}(v)=\exp\!\bigg(\!\int_0^{\varepsilon}\!\!\!ds\,\bar{A}_a^i\tau^i\dot{l}^a_{\nu}(s)\!\bigg)
=e^{c^i\tau^i}
\end{align}
and the related link excitation states, analogous to \eqref{shadow_basis}, takes the form
\begin{align}
\label{red_gravity}
\big|m^{(i)}_v\@\big>:=\exp\!\bigg[\i m^{(i)}_v c^{(i)}\bigg]\,,
\quad
2m^{(i)}\in\mathds{Z}\,.
\end{align}
Then, the actions of the lattice-regularized variables on these states are
\begin{align}
\label{red_action_c}
\hat{c}^{(i)}\big|m^{(i)}_v\@\big>:=&\,-\frac{2}{\varepsilon}\,\text{tr}\big(\tau^{(i)}h^{(i)}\big)\big|m^{(i)}_v\@\big>
=\frac{\i}{\varepsilon}\Bigg(\bigg|m^{(i)}_v-\frac{1}{2}\bigg>-\bigg|m^{(i)}_v+\frac{1}{2}\bigg>\@\Bigg)
\intertext{and}
\label{red_action_p}
\hat{p}^{(i)}\big|m^{(i)}_v\@\big>:=&\,-\i\kbar\frac{\partial}{\partial c^{(i)}\@}\big|m^{(i)}_v\@\big>
=m^{(i)}_v\kbar\big|m^{(i)}_v\@\big>\,.
\end{align}
The parameter $m^{(i)}_v$ is linked to the spin number $j^{(i)}_v$ by the relation $j^{(i)}_v=\big|m^{(i)}_v\big|$.

The last step toward formulation of the node-symmetric toy-model states on which the structure of the cosmological sector of CQGR will be tested is needed. To indicate the basic cell states centered at nodes (see FIG.~\ref{cellNORM}), one has to split the link states initially formulated to describe the states for LQC (see FIG.~\ref{spiderLQC}). This fitting of the well-known LQC shadow states to the analysis in this article is specified in the following relation:
\begin{align}
\label{red_link}
\big|m^{(i)}_v\@\big>=\exp\!\Big[\i\vec{m}^{(i)}_{v\mathstrut}c^{(i)}\Big]\exp\!\Big[\i\cev{m}^{(i)}_{\@v+\varepsilon^{(i)}}c^{(i)}\Big]
=\Big|\vec{m}^{(i)}_{v\mathstrut}\@\Big>\otimes\Big|\cev{m}^{(i)}_{\@v+\varepsilon^{(i)}}\@\Big>
\,,
\end{align}
where the oriented link $l^{(i)}\@(v)$ that starts at point $v$ was split in half,
\begin{align}
l^{(i)}\@(v)=\vec{l}{\.}^{(i)}\@(v)\Big[\,\cev{l}{}^{(i)}\!\big(v+\varepsilon^{(i)}\big)\@\Big]^{\@-1}\,.
\end{align}
The quantity $v\mp\varepsilon^{(i)}$ labels the nearest node along the negatively/positively-oriented $i$-th direction. In this way, two paths, $\vec{l}{\.}^{(i)}\@(v)$ and $\big[\:\cev{l}{}^{(i)}\!\big(v+\varepsilon^{(i)}\big)\@\big]^{-1}$, which have the following properties:
\begin{align}
\begin{split}
l^{(i)}\@(v)(0)&=\vec{l}{\.}^{(i)}\@(v)(0)=\Big[\,\cev{l}{\.}^{(i)}\@(v)\@\Big]^{\@-1}\!(0)\,,
\\
l^{(i)}\@(v)(1/2)&=\vec{l}{\.}^{(i)}\@(v)(1)=\Big[\,\cev{l}{\.}^{(i)}\!\big(v+\varepsilon^{(i)}\big)\@\Big]^{\@-1}\!(0)\,,
\\
l^{(i)}\@(v)(1)&=\Big[\,\cev{l}{\.}^{(i)}\!\big(v+\varepsilon^{(i)}\big)\@\Big]^{\@-1}\!(1)
=\vec{l}{\.}^{(i)}\!\big(v+\varepsilon^{(i)}\big)(0)\,,
\end{split}
\end{align}
were created. Then, the quantum numbers became fitted to this structure by postulating the simple averaging $\vec{m}^{(i)}_{v\mathstrut}=\cev{m}^{(i)}_{\@v+\varepsilon^{(i)}}=\frac{1}{2}m^{(i)}_v$\footnote{It is worth noting that this arithmetical mean corresponds to the averaging of the division of the analogous gravitational momentum, which would be constructed by the correspondence principle related to the original shadow states.}. This completes the definition of the node-centered states that share the symmetry of both the volume operator and the scalar field distribution. These simple toy-model states are
\begin{align}
\label{link_reduced}
\Big|\bar{m}^{(i)}_v\@\Big>:
=\Big|\frac{1}{2}m^{(i)}_{\@v-\varepsilon^{(i)}},\frac{1}{2}m^{(i)}_v\@\Big>
=\Big|\cev{m}^{(i)}_{v\mathstrut}\@\Big>\otimes\Big|\vec{m}^{(i)}_{v\mathstrut}\@\Big>
=\exp\!\bigg[\frac{\i}{2}\Big(m^{(i)}_{\@v-\varepsilon^{(i)}}+m^{(i)}_{v\mathstrut}\Big)c^{(i)}\bigg]
=\exp\!\Big[\i\bar{m}^{(i)}_{v\mathstrut}c^{(i)}\Big]\,.
\end{align}
The lattice-regularized canonical variables have the following actions on these basis states,
\begin{align}
\label{action_c_n}
\hat{c}^{(i)}\big|\bar{m}^{(i)}_v\@\big>=&\,-\frac{2}{\varepsilon}\,\text{tr}\bigg(\tau^{(i)}h^{(i)}_\frac{1}{2}\bigg)
\Big(\Big|\cev{m}^{(i)}_{v\mathstrut}\@\Big>\otimes\Big|\vec{m}^{(i)}_{v\mathstrut}\@\Big>\Big)
=\frac{\i}{\varepsilon}\bigg(\Big|\bar{m}^{(i)}_{\@v-\varepsilon^{(i)}\!}-\frac{\varepsilon}{2}\Big>-\Big|\bar{m}^{(i)}_v+\frac{\varepsilon}{2}\Big>\!\bigg)
\intertext{and}
\label{action_p_n}
\hat{p}^{(i)}\big|\bar{m}^{(i)}_v\@\big>=&\;\bar{m}^{(i)}_v\kbar\big|\bar{m}^{(i)}_v\@\big>\,.
\end{align}
Notice that in the former equation in \eqref{action_c_n}, the half-link-adjusted holonomy operator is
\begin{align}
h^{(i)}=e^{\varepsilon A^{(i)}_{(a)}{}^{0\!}e^{(i)}_{(a)}\tau^{(i)}}
\to
\ h^{(i)}_\frac{1}{2}:=e^{\frac{\varepsilon}{2} A^{(i)}_{(a)}{}^{0\!}e^{(i)}_{(a)}\tau^{(i)}}\!=e^{\frac{1}{2}c^i\tau^i}.
\end{align}

Then the coherent states analogous to \eqref{coherent_gravity} are given by the formula
\begin{align}
\label{coherent_QRLG}
\big|\ \big>^{\!(\text{gr}\@)}:=
\sum_v\bigotimes_i^3\bigg[
\Big(
\big<\vec{\mathfrak{c}}_{\@v\!}^{\.(\@i\@)\@\@}(A)\big|\vec{\mathfrak{c}}_{\@v\!}^{\.(\@i\@)\@\@}(A)\big>
\Big)^{\@\@-\frac{1}{2}}
\big|\vec{\mathfrak{c}}_{\@v\!}^{\.(\@i\@)\@\@}(A)\big>
\otimes
\Big(
\big<\,\cev{\!\mathfrak{c}}_{\@v\!}^{(\@i\@)\@\@}(A)\big|\,\cev{\!\mathfrak{c}}_{\@v\!}^{(\@i\@)\@\@}(A)\big>
\Big)^{\@\@-\frac{1}{2}}
\big|\,\cev{\!\mathfrak{c}}_{\@v\!}^{(\@i\@)\@\@}(A)\big>
\bigg]\,.
\end{align}
The corresponding link-oriented shadow coherent state is
\begin{align}
\label{shadow_QRLG}
\big|\.\cev{\@\vec{\.\reflectbox{\ensuremath{\mathfrak{c}}}\@}\.}_{\@\@v\!}^{\@(\@i\@)\@\@}\@(A)\big>:=
\sum_{\.\cev{\@\vec{\.\reflectbox{\scriptsize \ensuremath{m}}\@}\.}\@^{(i)}_v\!}
\exp\!\Bigg[
{-\frac{1}{2d^2}\bigg(\.\cev{\@\vec{\.\reflectbox{\ensuremath{m}}\@}\.}\@^{(i)}_v\@-\frac{p^{(i)}\@}{\kbar}\bigg)^{\!\!2}}
\Bigg]
\exp\!\Bigg[
{-\i\bigg(\.\cev{\@\vec{\.\reflectbox{\ensuremath{m}}\@}\.}\@^{(i)}_v\@-\frac{p^{(i)}\@}{\kbar}\bigg)}c^{(i)}
\Bigg]\big|\.\cev{\@\vec{\.\reflectbox{\ensuremath{m}}\@}\.}\@^{(i)}_v\@\big>\,.
\end{align}
The node-centered coherent states, adjusted to \eqref{link_reduced} are defined analogously,
\begin{align}
\label{coherent_normalized}
\!\bar{\,\big|\ \big>}^{\!(\text{gr}\@)}:=
\sum_v\bigotimes_i^3\Big(
\big<\bar{\mathfrak{c}}_{\@v\!}^{(\@i\@)\@\@}(A)\big|\bar{\mathfrak{c}}_{\@v\!}^{(\@i\@)\@\@}(A)\big>
\Big)^{\@\@-\frac{1}{2}}
\big|\bar{\mathfrak{c}}_{\@v\!}^{(\@i\@)\@\@}(A)\big>\,,
\end{align}
where their node-centered shadow state coefficients are
\begin{align}
\label{shadow_normalized}
\big|\bar{\mathfrak{c}}_{\@v\!}^{(\@i\@)\@\@}(A)\big>:=
\sum_{\bar{m}^{(i)}_v}
\exp\!\Bigg[
{-\frac{1}{2d^2}\bigg(\bar{m}^{(i)}_v\@-\frac{p^{(i)}\@}{\kbar}\bigg)^{\!\!2}}
\Bigg]
\exp\!\Bigg[
{-\i\bigg(\bar{m}^{(i)}_v\@-\frac{p^{(i)}\@}{\kbar}\bigg)}c^{(i)}
\Bigg]\big|\bar{m}^{(i)}_v\@\big>\,.
\end{align}

It is worth noting that these states (and the ones in \eqref{shadow_gravity}, before the symmetrization) satisfy the coherent states requirements discussed concerning different aspects of LQG \cite{Thiemann:2000bw,Thiemann:2000ca,Thiemann:2000bx,Ashtekar:2002sn}. The detailed analysis of the constructions of analogous states as the gauge-invariant projection of a product over links of heat kernels for the complexification of group elements can be found in \cite{Alesci:2014uha}.

Finally, the reader more familiar with LQC might be interested in whether the simplified node-symmetrized model leads to the same expectation values of the canonical operators. By deriving the expectation value of the $\hat{c}^{(i)}$ operator,
\begin{align}
\begin{split}
\big<\bar{\mathfrak{c}}_{\@v\!}^{\@(\@i\@)\@\@}(A)\big|\hat{c}^{(i)}\big|\bar{\mathfrak{c}}_{\@v\!}^{\@(\@i\@)\@\@}(A)\big>
=&\;
\big._{_R\!\!}\big<\,\cev{\!\mathfrak{c}}_{\@v\!}^{(\@i\@)\@\@}(A)\big|
\otimes
\big._{_R\!\!}\big<\vec{\mathfrak{c}}_{\@v\!}^{\.(\@i\@)\@\@}(A)\big|
\hat{c}^{(i)}
\big|\vec{\mathfrak{c}}_{\@v\!}^{\.(\@i\@)\@\@}(A)\big>_{\!\!_R}
\otimes
\big|\,\cev{\!\mathfrak{c}}_{\@v\!}^{(\@i\@)\@\@}(A)\big>_{\!\!_R}
\\
=&\;
\frac{2}{\varepsilon}\exp\!\Bigg[\@-\@\bigg(\frac{\varepsilon}{2d}\bigg)^{\!\!2}\Bigg]
\sum_{m^{(i)}_v\!}\exp\!\Bigg[\@-\frac{1}{d^2}\bigg(\frac{p^{(i)}\@}{\kbar}-\bar{m}^{(i)}_v\@\bigg)^{\!\!2}\Bigg]
\sin\bigg[\frac{\varepsilon}{2}c^{(i)}+\i\varepsilon\bigg(\frac{p^{(i)}\@}{\kbar}-\bar{m}^{(i)}_v\@\bigg)\bigg]\,,
\end{split}
\end{align}
one obtains the result analogous to the one know for LQC. The identification would be exact after the replacement
\begin{align}
\label{substitution}
\bar{m}^{(i)}_v\to\frac{1}{2}\mu^{(i)}_v\,.
\end{align}
By substituting the appropriate correspondence principle
\begin{align}
\label{correspondence}
\bar{m}^{(i)}_v\to\frac{p^{(i)}\@}{\kbar}\,,
\end{align}
the result can be recast in the simple form
\begin{align}
\big<\hat{c}^{(i)}\big>
=c^{(i)}\Big(1+\mathcal{O}\big(\varepsilon^2\big)\Big)\,.
\end{align}
Analogously, the expectation value of the reduced flux operator becomes
\begin{align}
\big<\hat{p}^{(i)}\big>
=p^{(i)}\,.
\end{align}
The last pair of equations will be enough to discuss the SE quantum matter coupling to LQG concerning the BI of the related semiclassical results.


\section{Quantum corrections}\label{IV}

\subsection{Ehrenfest theorem and Heisenberg equation}\label{IV.1}

\noindent
In this section the matrix elements on the coherent states of what could be the cosmological reduction of CQGR are analyzed. All the conclusions are going to be studied in the formalism general enough to be directly related with LQC, QRLG, CCGR, and analogous models. 

The semiclassical dynamics of the whole cosmological system is given by the Heisenberg equations
\begin{align}
\label{Ehrenfest_c}
\frac{d\big<\hat{c}\big>^{\@(\text{gr}\@)}\!\!\!\!}{dt}\ 
=&\;\frac{1}{\i\hbar}\bigg(
\Big<\big[\hat{c},\hat{H}^{(\text{gr}\@)}\big]\Big>^{\@\@(\text{gr}\@)}
\!+\Big<\big[\hat{c},\hat{H}^{(\text{matt})}\big]\Big>^{\@\@(\text{gr}\@)}
\bigg)
=\frac{1}{\i\hbar}
\Big<\big[\hat{c},\hat{H}^{(\text{gr}\@)}\big]\Big>^{\@\@(\text{gr}\@)}
\!+\Delta_c^{H^{(\text{matt})}}\,,
\\
\label{Ehrenfest_p}
\frac{dp}{dt}
=&\;\frac{1}{\i\hbar}\Big<\big[\hat{p},\hat{H}^{(\text{gr}\@)}\big]\Big>^{\@\@(\text{gr}\@)}\,,
\\
\label{Ehrenfest_Phi}
\frac{d\big<\hat{\textbf{\straightphi}}\big>^{\@(\text{matt})}\!\!\!\!}{dt}\ 
=&\;\frac{1}{\i\hbar}
\bigg<
\Big<\big[\hat{\textbf{\straightphi}},\hat{H}^{(\text{matt})}\big]\Big>^{\@\@(\text{matt})}
\bigg>^{\!\!(\text{gr}\@)}
=\frac{1}{\i\hbar}
\Big<\big[\hat{\textbf{\straightphi}},\hat{H}^{(\text{matt})}\big]\Big>^{\@\@(\text{matt})}
\!+\Delta_{\textbf{\straightphi}}^{|\,\rangle^{\@(\text{gr}\@)}}\,.
\end{align}
Analogously to \eqref{coherent_tensor}, the symbol $\textbf{\straightphi}$ represents any matter field. Precisely, it denotes either the canonical field variable or the corresponding conjugate momentum. The indices labeling the directions and position of operators were omitted for simplicity. It was also assumed that the gravitational and matter fields are not explicitly time dependent and their evolution is encoded only in the equations of motion.

The \uline{quantum GR corrections} both in \eqref{Ehrenfest_c} and \eqref{Ehrenfest_Phi} are of the same order in the inverse spin number, precisely
\begin{align}
\label{corrections1}
\Delta_c^{H^{(\text{matt})}}\!\propto\Delta_{\textbf{\straightphi}}^{|\,\rangle^{\@(\text{gr}\@)}}\!\propto\frac{1}{\bar{m}^2}\,.
\end{align}
Here, the large quantum number approximation $|\bar{m}|\gg1$ is needed\footnote{This approximation relates the single fiducial cell formula with the Hamiltonian on $\Sigma_t$ in the continuum limit \cite{Thiemann:2007zz,Dittrich:2014ala}. Generalization to any value of $\bar{m}$ is possible, but it would require a redefinition of the coherent states. Heuristically, this could be done replacing $\bar{m}$ with $\bar{m}_{\varepsilon}:=\bar{m}/\varepsilon$ in the definition \eqref{link_reduced}. Precise approach would require redefinition of LQG and the appropriate phase space reduction. The first articles concerning the former issue was recently announced, \textit{cf.} \cite{Bilski:2020poi}. The regulator-independent formulation of the lattice reduced theory is based on \cite{Bilski:2020xfq} and is given in \cite{Bilski:2021_LCC}.}.
The detailed derivation and the exact numerical value of these SE-sourced corrections is model dependent --- see for instance \cite{Bilski:2016pib,Hossain:2010wy,Alesci:2014uha}. In what follows, only the structure of the gravitational degrees of freedom contributing to the corrections from equation \eqref{Ehrenfest_Phi} is going to be used in the general covariance verification. Consequently, the decomposition into the gravity- and matter-related expressions will be later introduced in \eqref{corrections_decomposition}. Finally, the explicit derivation of the structure of the matter sector-related GR corrections, needed for the verification procedure, will be given in Sec.~\ref{IV.3}.

The second type of the classical dynamics perturbations that come from the \uline{quantum gravitational corrections} will be denoted by $\delta_{\dot{c}}$, $\delta_c$, and $\delta_p$. These quantities are sourced by the terms
\begin{align}
\label{term0}
\frac{d\big<\hat{c}\big>^{\@(\text{gr}\@)}\!\!\!\!}{dt}\ 
=&\;\frac{dc}{dt}\big(1+\delta_{\dot{c}}\big)\,,
\\
\label{term1}
\Big<\big[\hat{c},\hat{H}^{(\text{gr}\@)}\big]\Big>^{\@\@(\text{gr}\@)}\!
=&\;\i\hbar\frac{\delta H^{(\text{gr}\@)}\!\!}{\delta p}
\big(1+\delta_c\big)\,,
\intertext{and}
\label{term2}
\Big<\big[\hat{p},\hat{H}^{(\text{gr}\@)}\big]\Big>^{\@\@(\text{gr}\@)}\!
=&\;-\i\hbar\frac{\delta H^{(\text{gr}\@)}\!\!}{\delta c}
\big(1+\delta_p\big)\,,
\end{align}
respectively. They have a qualitatively different structure than the quantum GR corrections in \eqref{corrections1}, by satisfying
\begin{align}
\label{corrections2}
\delta_{\dot{c}}\propto\delta_c\propto\delta_p\propto\varepsilon^2\,.
\end{align}
Another difference between these corrections is in the fact that the gravitational corrections are functionals of the connection: $\delta_{\dot{c}}=\delta_{\dot{c}}[c]$, $\delta_c=\delta_c[c]$, and $\delta_p=\delta_p[c]$ and are related only to the regularization of the gravitational sector. The GR corrections are related to the SE restriction imposition and depend only on quantum numbers. Thus, by the correspondence principle in \eqref{correspondence}, they indirectly depend on the reduced flux, which is directly related to the spatial metric tensor. However, they are independent of the gravitational correction. This feature can be written as
\begin{align}
\label{dependence}
\frac{\partial}{\partial c}\.\Delta_c^{H^{(\text{matt})}}
=
\frac{\partial}{\partial c}\.\Delta_{\textbf{\straightphi}}^{|\,\rangle^{\@(\text{gr}\@)}}
=0\,.
\end{align}
Notice also that by neglecting the evolution of the gravitational degrees of freedom in \eqref{Ehrenfest_Phi}, this Heisenberg equation takes the form
\begin{align}
\label{Ehrenfest_special}
\frac{d\big<\hat{\textbf{\straightphi}}\big>^{\@(\text{matt})}\!\!\!\!}{dt}\ 
=\frac{1}{\i\hbar}
\Big<\big[\hat{\textbf{\straightphi}},\hat{H}^{(\text{matt})}\big]\Big>^{\@\@(\text{matt})},
\quad
\Delta_{\textbf{\straightphi}}^{|\,\rangle^{\@(\text{gr}\@)}}=0\,.
\end{align}
Thus it is the same as the Heisenberg equation for the lattice-regularized QFT on curved spacetime. It is worth mentioning that if one included the effects of this evolution, one would obtain additional dynamical corrections of order $\bar{m}^2/p^3$ \cite{Hossain:2010wy}.

The structure of the quantum GR corrections in \eqref{corrections1} is the essential element for the analysis in this article. These are the only gravitational degrees of freedom-dependent corrections present in the matter sector. They were sourced by SE and their structure verifies BI of this system. Therefore general covariance can be tested investigating these quantum perturbations structure.

The gravitational coupling in the matter sector of GR is implemented by the multiplication by the $q^{\pm1/2}$ factors and/or by the contraction with the $q_{ab}$ metric tensor --- see the expressions in \eqref{YMscalar} and \eqref{scalar_scalar}. After the lattice regularization, these recalled expressions take the forms given in formulas \eqref{H_vect} and \eqref{H_scal}, respectively. The GR corrections in these formulas will be sourced by the quantized version of the terms
\begin{align}
\label{gr_coupling}
\,\text{tr}\Big(\@\tau^ih_{\@(\@a\@)}^{-1}\@(v)\big\{\mathbf{V}^n\@(v),h_{\@(\@a\@)}^{\mathstrut}\@(v)\big\}\@\Big)\,,
\quad
n\in\mathds{Q}_+
\end{align}
that were constructed by using the relation in \eqref{E_trick}. At the quantum level, the aforementioned quantity becomes the trace of the product of the $\mathfrak{su}(2)$ generator and the operators in \eqref{inverse_general}, and its structure varies for different matter fields. Moreover, even in the Hamiltonian constraint for a given field, the elements with different power of volume in \eqref{gr_coupling} are present --- compare \eqref{H_scal_mom}, \eqref{H_scal_der}, and \eqref{H_scal_pot}. To study these differences, the expression in \eqref{Ehrenfest_Phi} has to be decomposed more specifically.

One should first observe the following relation,
\begin{align}
\label{Ehrenfest_relation}
\bigg<
\Big<\big[\hat{\textbf{\straightphi}},\hat{H}^{(\text{matt})}\big]\Big>^{\@\@(\text{matt})}
\bigg>^{\!\!(\text{gr}\@)}\!
=
\bigg<
\Big[\hat{\textbf{\straightphi}},\big<\hat{H}^{(\text{matt})}\big>^{\@(\text{gr})}\Big]
\bigg>^{\!\!(\text{matt}\@)}.
\end{align}
This leads to the conclusion that the structure of $\Delta_{\textbf{\straightphi}}^{|\,\rangle^{\@(\text{gr}\@)}}\@$ depends only on the matrix element $\big<\hat{H}^{(\text{matt})}\big>^{\@(\text{gr})}$. Then, by splitting the matter sector Hamiltonian into the contributions from different fields $\textbf{\straightphi}_{\@\alpha}$\footnote{In this general approach , $\textbf{\straightphi}_{\@\alpha}$ represents any matter field. In the case of the simplified cosmological model with bosonic matter and with the Hilbert space given in \eqref{total_space}, only two different matter fields are considered: $\textbf{\straightphi}_{\und{A}}:=\und{A}$ and $\textbf{\straightphi}_{\varphi}:=\varphi$.}, one finds the following decomposition:
\begin{align}
\label{semi_matt}
H^{(\text{matt})}\@=\sum_{\alpha}H^{(\@\textbf{\straightphi}_{\@\alpha}\@)}\@
=\sum_{\alpha}\Big(H^{(\@\textbf{\straightphi}_{\@\alpha}\@)}_{_\text{one}\mathstrut\!}\@+H^{(\@\textbf{\straightphi}_{\@\alpha}\@)}_{_\text{two}\mathstrut\!}\@+...\Big)
=:\sum_{\alpha}\bigg(\sum_{^\text{elements}}\!\!H^{(\@\textbf{\straightphi}_{\@\alpha}\@)}_{_\text{element}\mathstrut\!}\bigg)\,.
\end{align}
The second splitting in the formula above is given by the introduction of the terms $H^{(\@\textbf{\straightphi}_{\@\alpha}\@)}_{_\text{one}\mathstrut\!}\@,H^{(\@\textbf{\straightphi}_{\@\alpha}\@)}_{_\text{two}\mathstrut\!}\@,...$ that label different elements in the $\textbf{\straightphi}_{\@\alpha}$ field Hamiltonian. For instance, the Hamiltonian of $\textbf{\straightphi}_{\und{A}}$ decomposes as follows: $H^{(\und{A})}=:H^{(\@\textbf{\straightphi}_{\@\und{A}}\@)}=H^{(\@\textbf{\straightphi}_{\@\und{A}}\@)\@}_{_\text{elec}\mathstrut\!}+H^{(\@\textbf{\straightphi}_{\@\und{A}}\@)}_{_\text{magn}\mathstrut\!}\.$.

The matrix element derivation is a linear operation, thus without loss of generality it is enough to focus on a single element
\begin{align}
\label{matrix_element}
\Big<\hat{H}^{(\@\textbf{\straightphi}_{\@\alpha}\@)}_{_\text{element}\mathstrut\!}\Big>^{\@\@(\text{gr}\@)}\!
=H^{(\@\textbf{\straightphi}_{\@\alpha}\@)}_{_\text{element}\mathstrut\!}
\Big(
1+\delta^{(\@\textbf{\straightphi}_{\@\alpha}\@)}_{_\text{element}\mathstrut\!}\@
+\delta'^{(\@\textbf{\straightphi}_{\@\alpha}\@)}_{_\text{element}\mathstrut\!}\@
+...
\Big),
\end{align}
where $\delta^{(\@\textbf{\straightphi}_{\@\alpha}\@)}_{_\text{element}\mathstrut\!}\@\propto1/\bar{m}^2$, $\delta'^{(\@\textbf{\straightphi}_{\@\alpha}\@)}_{_\text{element}\mathstrut\!}\@\propto1/\bar{m}^4$, \textit{etc}. For simplicity, the terms of order $1/\bar{m}^4$ and smaller are going to be neglected. Consequently, the quantum GR corrections to the matter sector are expressible by
\begin{align}
\label{corrections_decomposition}
\Delta_{\textbf{\straightphi}}^{|\,\rangle^{\@(\text{gr}\@)}}
=\frac{1}{\i\hbar}\sum_{\alpha}\sum_{^\text{elements}}\!\!
\bigg<
\Big[\hat{\textbf{\straightphi}}_{\@\alpha},\hat{H}^{(\@\textbf{\straightphi}_{\@\alpha}\@)}_{_\text{element}\mathstrut\!}\Big]
\bigg>^{\@\@(\text{matt})}
\delta^{(\@\textbf{\straightphi}_{\@\alpha}\@)}_{_\text{element}\mathstrut\!}\,,
\end{align}
where the linearity of a commutator was used. Finally, it should be pointed out that when the correspondence principle in \eqref{correspondence} is applied, the corrections become explicitly dreibein dependent, thus also metric tensor dependent. In the case of the vector field in the cosmological framework, the structure of this dependence is readable from the expression
\begin{align}
\label{corrections_Hamiltonian_vector}
\bigg<
\Big<\big[\hat{\textbf{\straightphi}},\hat{H}^{(\underline{A})}\big]\Big>^{\@\@(\underline{A})}
\bigg>^{\!\!(\text{gr}\@)}\!
=
\sum_a^3
\bigg<
\Big<\big[\hat{\textbf{\straightphi}},\hat{H}^{(\underline{A})}_{\@(a)}\big]\Big>^{\@\@(\underline{A})}
\bigg>^{\!\!(\text{gr}\@)}\!
=
\sum_a^3
\bigg<
\Big[\hat{\textbf{\straightphi}},\hat{H}^{(\underline{A})}_{\@(a)\mathstrut\!}\Big]
\bigg>^{\@\@(\underline{A})}\!
\Big(
1+\delta^{(\underline{A})}_{\@(a)\mathstrut\!}
\Big)\,,
\end{align}
where
\begin{align}
\label{corrections_explicit_vector}
\delta^{(\underline{A})}_{\@(a)\mathstrut\!}
\propto
\frac{\kbar^2}{\big(p^{(a)}\big)^{\@\@2}}\,.
\end{align}
The form of the preceding outcome reflects the symmetry between the regularized elements in the Hamiltonian contributions in \eqref{H_vect_elec} and \eqref{H_vect_magn}. The analogous matrix element of the scalar field leads to the result
\begin{align}
\label{corrections_Hamiltonian_scalar}
\begin{split}
\bigg<
\Big<\big[\hat{\textbf{\straightphi}},\hat{H}^{(\varphi)}\big]\Big>^{\@\@(\varphi)}
\bigg>^{\!\!(\text{gr}\@)}\!
=
\bigg<
\Big[\hat{\textbf{\straightphi}},\hat{H}^{(\varphi)}_{\@_\text{mom}\mathstrut\!}\Big]
\bigg>^{\@\@(\varphi)}\!
\Big(
1+\delta^{(\varphi)}_{\@_\text{mom}\mathstrut\!}
\Big)+
\sum_a^3
\bigg<
\Big[\hat{\textbf{\straightphi}},\hat{H}^{(\varphi)}_{\@(a)\._\text{der}\mathstrut\!}\Big]
\bigg>^{\@\@(\varphi)}\!
\Big(
1+\delta^{(\varphi)}_{\@(a)\._\text{der}\mathstrut\!}
\Big)
+
\bigg<
\Big[\hat{\textbf{\straightphi}},\hat{H}^{(\varphi)}_{\@_\text{pot}\mathstrut\!}\Big]
\bigg>^{\@\@(\varphi)},
\end{split}
\end{align}
where
\begin{align}
\label{corrections_explicit_mom}
\delta^{(\varphi)}_{\@_\text{mom}\mathstrut\!}
&\propto
\sum_a^3\frac{\kbar^2}{\big(p^{(a)}\big)^{\@\@2}}\,,
\\
\label{corrections_explicit_der}
\delta^{(\varphi)}_{\@(a)\._\text{der}\mathstrut\!}
&\propto
\sum_{b\neq a}\frac{\kbar^2}{\big(p^{(b)}\big)^{\@\@2}}\,.
\end{align}
This outcome is not symmetric with respect to the metric tensor structure, hence the BI of the indicated quantum GR corrections is explicitly broken.

\subsection{General quantum relativity}\label{IV.2}

\noindent
Before describing the general covariance breaking in more details by the use of inconsistently selected regularization methods, it is worth it to state more precisely what is the general postulate of relativity \cite{Einstein:1916vd} in the context of quantum physics.

This postulate was originally formulated as follows.
\begin{quote}
\textit{The general laws of nature are to be expressed by equations which hold good for all systems of co-ordinates, that is, are covariant with respect to any substitutions whatever (generally co-variant).}
\\
\dots For the sum of \textit{all} substitutions in any case includes those which correspond to all relative motions of three-dimensional systems of co-ordinates.\,\dots Moreover, the results of our measuring are nothing but verifications of such meetings of the material points of our measuring instruments with other material points, coincidences between the hands of a clock and points of the clock dial, and observed point-events happening at the same place at the same time. \cite{Einstein:1916vd}
\end{quote}
The postulate and its explanation (see also their earlier formulations in \cite{Einstein:1907,Einstein:1911vc,Einstein:1907iag}) consists of two restrictions on a physical theory. The theory has to have SE, \textit{i.e.} the equations must be equivalently expressed in all systems of coordinates. It also has to have BI, \textit{i.e.} the predictions of the theory must be invariant under any substitution of a reference frame.

It should be emphasized that SE does not require any independence of coordinates or a metric involving formalism. The SE model can be explicitly formulated in a particular system of coordinates, however, the choice of any other system has to lead to the equivalent formulation. Furthermore, BI does not assume any restriction on a formalism. The BI theory has to provide universal predictions independently of any particular characteristics of an observer.

In this article it is assumed that a hypothetical CQGR model satisfies SE and BI. This does not provide any \textit{a priori} reference to gravity. Whatever content is included in the theory, it has to have SE and BI. This clarifies what GR means in the context of CQGR. Now, the meaning of quantum should be specified. This term is going to be understood in the sense of the quantum mechanical formalism adjusted to each component of CQGR. In particular, this assumes the probabilistic distributional framework, noncommutative relations and the predictions formulated as expectation values. The description of these features is well understood in terms of interacting quantum fields. Therefore, CQGR is going to be understood as QFT that satisfies SE and BI. It is not difficult to realize how to impose these latter restrictions on QFT.

The SE formulation of QFT means that the formulation of interacting quantum fields is equivalent in any system of coordinates. Therefore the action of any operator on any state cannot, in reality, depend on any reference frame. However, an apparent dependence is not excluded. This means that operator equations cannot be localized at any fixed spacetime points and the scalar product cannot depend on the position in a Fock space. All the interactions must consist of the relations between certain characteristics of quantum operators and states and cannot depend on any fixed reference frames that would classify these characteristics. Simply speaking, the framework of QFT has to be generally relative.

It is easy to see that the phenomenological models discussed in Sec.~\ref{II.5} violate the SE requirement. This excludes these models from being a candidate for a cosmological limit of CQGR funded by using the LQG's framework. It is then surprising that there are several studies toward formulation of phenomenological predictions by using these effective models. These models directly break SE in their constructions, which are based on the LQG's formalism that was created to describe the QFT of gravity in the SE way. Thus, these mentioned phenomenological applications as the predictions that could explain reality are methodologically inconsistent.

The framework satisfying the SE requirement was introduced in Sec.~\ref{III.1}. In Sec.~\ref{IV.3} details of the BI verification are explained. In general, the BI formulation of QFT means that the predictions of the theory are independent of any reference frame. In the context of quantum physics this restriction is directly related to the notion of observables. These are the indirectly measurable quantities in QFT. However, the predictions are not formulated in terms of observables. Only the eigenvalues of these operators are directly measurable. Therefore the BI of a quantum theory means BI spectra of all observables. These are the only quantities in which an observer verifies the laws of nature. They are described relatively if their predictions are formulated independently of how, when, and where they can be tested.

\begin{figure}[h]
\begin{center}
\begin{tikzpicture}[scale=0.75]
\draw[l,->]{(-4.25,-4.1) -- (-4.75,-4.9)};
\draw[ll]{(-4.25,1.9) -- (1.75,1.9)};
\draw[ll]{(-4.25,-4.1) -- (1.75,-4.1)};
\draw[ll]{(-4.25,1.9) -- (-4.25,-4.1)};
\draw[ll]{(1.75,1.9) -- (1.75,-4.1)};
\draw[ll]{(-4.25,1.9) -- (-3.25,3.5)};
\draw[ll]{(-4.25,-4.1) -- (-3.25,-2.5)};
\draw[ll]{(1.75,1.9) -- (2.75,3.5)};
\draw[ll]{(1.75,-4.1) -- (2.75,-2.5)};
\draw[lll]{(-3.25,3.5) -- (2.75,3.5)};
\draw[lll,->]{(-3.25,-2.5) -- (4.75,-2.5)};
\draw[lll,->]{(-3.25,-2.5) -- (-3.25,5.5)};
\draw[lll]{(2.75,3.5) -- (2.75,-2.5)};
\node at (-4.15,-4.9) {$1/c$};
\node at (4.5,-2.25) {$\hbar$};
\node at (-3.55,5.25) {$G$};
\node at (-1.5,-2.8) {Galilei-Newtonian};
\node at (-1.5,-3.2) {theory};
\node at (4.4,-2.8) {non-relativistic};
\node at (4.4,-3.2) {quantum mechanics};
\node at (-1.3,3.2) {classical mechanics};
\node at (-1.3,2.8) {(Newtonian gravitation)};
\node at (4.4,3.2) {non-relativistic};
\node at (4.4,2.8) {quantum theory};
\node at (-2.5,-4.4) {special relativity};
\node at (3.3,-4.4) {quantum field};
\node at (3.3,-4.75) {theory};
\node at (-2.48,1.6) {general relativity};
\node at (3.52,1.6) {quantum};
\node at (3.52,1.2) {general relativity};
\end{tikzpicture}
\vspace{-10pt}
\caption{Bronstein cube in $c^{-1}\hbar\.G$ orientation}
\label{cube}
\vspace{-5pt}
\end{center}
\end{figure}
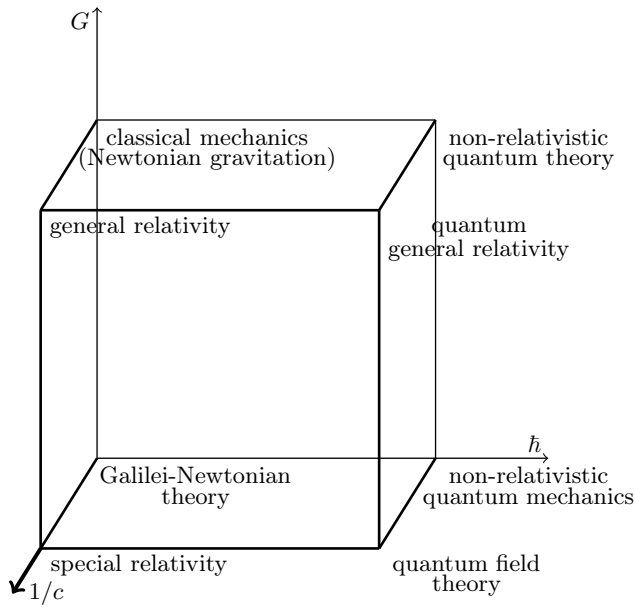
Eventually, the model of quantum GR (see FIG.~\ref{cube}) that is based on the framework of LQG assumes canonical quantization procedure. This last property defines how the classical and quantum descriptions are related. As pointed out in \cite{Bilski:2020poi}, this relation is not clear in LQG, which is a separate problem and is not going to be discussed in this article. In general, the canonical quantization is defined as a replacement of canonical variables with their operator representations and a replacement of Poisson brackets with commutators. Then, the change of variables into operators should preserve the relative orientations and positions of these objects, accordingly to all the frames indicated in a theory (this is not properly implemented in the original canonical formulation of LQG \cite{Thiemann:1996aw,Thiemann:2007zz}, \textit{cf.} \cite{Bilski:2020poi}). In this way, the gauge invariance is properly preserved both locally and globally.

Concluding, the properly formulated candidate for CQGR has to satisfy the following restrictions.
\begin{enumerate}[a)]
\item
\label{rA}
Quantization is performed in a canonical procedure that preserves gauge invariance both locally and globally.
\item
\label{rB}
Equivalence principle is satisfied strongly for all the fields and all the coordinates systems.
\item
\label{rC}
All the predictions are background independent, thus are the same for any observer.
\end{enumerate}
The first condition allows one to analyze phase space-reduced versions of CQGR and obtain results related to the general model. The second restriction is imposed already in the construction of theories, hence one can focus only on the models that satisfy this condition. To verify the last restriction, kinematics of the theory has to be derived and its dynamics has to be formulated.

Finally, it is worth it to emphasize that besides the theoretical notion of the analysis in this article, one can also indicate its practical value. In order to describe the physical meaning of the correct formulation of CQGR cosmological reduction, an independent consideration of the matrix element of the HCO only with respect to the matter or to the gravitational degrees of freedom is going to be discussed. This will be presented on the Bronstein cube \cite{Bronstein_cube} illustrative diagram in FIG.~\ref{cube}.

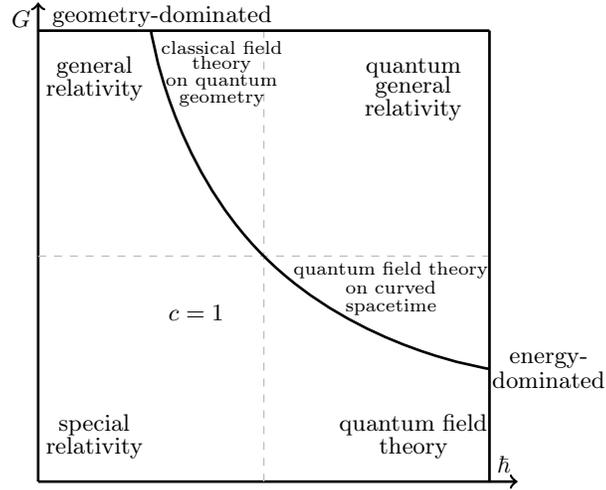
\begin{figure}[h]
\begin{center}
\begin{tikzpicture}[scale=0.75]
\draw[c,->]{(-4,-4) -- (-4,4.5)};
\draw[c,->]{(-4,-4) -- (4.5,-4)};
\draw[d]{(-4,0) -- (4,0)};
\draw[d]{(0,-4) -- (0,4)};
\draw[c]{(-4,4) -- (4,4)};
\draw[c]{(4,-4) -- (4,4)};
\draw[c]plot[smooth, tension=1] coordinates 
{(-2,4) (0,0) (4,-2)};
\node at (-1.2,-1) {$c=1$};
\node at (4.25,-3.7) {$\hbar$};
\node at (5.05,-1.85) {energy-};
\node at (5.05,-2.2) {dominated};
\node at (-4.3,4.2) {$G$};
\node at (-1.8,4.25) {geometry-dominated};
\node at (-3.0,3.35) {general};
\node at (-3.0,2.95) {relativity};
\node at (-0.75,3.7) {\scriptsize classical field};
\node at (-0.75,3.4) {\scriptsize theory};
\node at (-0.75,3.1) {\scriptsize on quantum};
\node at (-0.75,2.8) {\scriptsize geometry};
\node at (2.65,3.35) {quantum};
\node at (2.65,3.0) {general};
\node at (2.65,2.6) {relativity};
\node at (2.25,-0.25) {\scriptsize quantum field theory};
\node at (2.25,-0.55) {\scriptsize on curved};
\node at (2.25,-0.9) {\scriptsize spacetime};
\node at (-3.0,-3.0) {special};
\node at (-3.0,-3.4) {relativity};
\node at (2.65,-3.0) {quantum field};
\node at (2.65,-3.4) {theory};
\end{tikzpicture}
\caption{Planck hyperbola, $l_P=\text{constant}$}
\label{hyperbola}
\vspace{-5pt}
\end{center}
\end{figure}

The semiclassical low energy approximation of quantum matter excitations corresponds to the classical matter fields theory on quantum geometry. The semiclassical slightly curved approximation of quantum geometry coincides with QFT on classical curved spacetime. The first approximation corresponds to taking the $\hbar\to0$ limit and the second one to taking the $G\to0$ limit. These approximations can be relevant in different physical processes --- see FIG.~\ref{hyperbola}.

The classical and flat limit of CQGR can be (not rigorously) understood as $l_P\to0$. To study only the semiclassical corrections of the theory, one can expand the results around a small (but not zero) value of the Planck length. This expansion can be represented by the hyperbola in FIG.~\ref{hyperbola}. This figure expresses the $c=1$ face of the Bronstein cube in FIG.~\ref{cube} and the sketched curve is going to be called the Planck hyperbola.

From the cosmological perspective, the well-known QFT on curved spacetime approach \cite{Birrell:1982ix} can be applied, for instance, to explain the details of the inflation process \cite{Guth:1980zm}. The classical field theory on quantum geometry can be used as an approximation of the early phase in the Universe evolution. This model predicts, for instance, a big bounce scenario at the origin of the Universe \cite{Bojowald:1999tr,Bojowald:2001xe}. Therefore, to understand the whole Universe evolution, even only approximately by studying the semiclassical corrections of CQGR, one needs to be able to smoothly move along the Planck hyperbola. Then, to be sure that this move is smooth, the cosmological reduction of CQGR has to be properly constructed. Although, if the general theory is not completely formulated, one cannot be absolutely sure that the cosmological limit of CQGR can be precisely applied as a model of the Universe. This argumentation provides the physical motivation for the verification of the covariant structure of the corrections indicated in Sec.~\ref{IV.1}.

\subsection{Violation of general covariance}\label{IV.3}

\noindent
The explicit form of the quantum GR corrections indicated in expressions \eqref{corrections_Hamiltonian_vector} and \eqref{corrections_Hamiltonian_scalar} depends on the state on which the operator constructed from the term in \eqref{gr_coupling} acts. The symmetrized shadow states in \eqref{shadow_normalized}, which satisfy basic properties of correctly formulated coherent states, are based on the links excitations states in \eqref{link_reduced}. Thus only the results of the reduced operators actions on these links excitations states have to be verified.

The holonomy operator in \eqref{red_holonomy} of the constant Abelian connection resulting from the phase space reduction of CQGR leads to the following simplification of the corrections generating operator,
\begin{align}
\label{gr_coupling_matter}
\,\text{tr}\bigg(\@\tau^i\hat{h}_{\@(\@a\@)}^{-1}\Big[\hat{\mathbf{V}}^n,\hat{h}_{\@(\@a\@)}^{\mathstrut}\Big]\@\bigg)
=
\sin\!\Bigg(\!\frac{\hat{c}_{(\@a\@)}^{\.i}\@}{2}\!\Bigg)\hat{\mathbf{V}}^n\cos\!\Bigg(\@\frac{\vec{\hat{c}}_{(\@a\@)}\@}{2}\@\Bigg)
-\cos\!\Bigg(\@\frac{\vec{\hat{c}}_{(\@a\@)}\@}{2}\@\Bigg)\hat{\mathbf{V}}^n\sin\!\Bigg(\!\frac{\hat{c}_{(\@a\@)}^{\.i}\@}{2}\!\Bigg)\,.
\end{align}
For clearness, the projection of the spatial directions into the internal ones and the related correction of variables' weights \cite{Bilski:2019tji,Ashtekar:2003hd} is not explicitly written in this expression. The vector symbol over the connection in $\vec{\hat{c}}_{(\@a\@)}$ indicates the direction-independent series representation of the cosine operator functional
\begin{align}
\label{cosine}
\cos\!\Bigg(\@\frac{\vec{\hat{c}}_{(\@a\@)}\@}{2}\@\Bigg)
=\sum_{k=0}^{\infty}\frac{(-1)^k}{(2k)!}\bigg(\@\frac{\hat{c}_{(\@a\@)}^{\.i}\hat{c}_{(\@a\@)}^{\.i}\@}{4}\@\bigg)^{\!\!k}\,.
\end{align}

By assuming only cuboidal cells, the action of the volume operator in \eqref{gr_coupling_matter} expressed in terms of the reduced momenta in \eqref{red_E} simplifies into the operator constructed from the following quantity:
\begin{align}
\label{volume}
\bar{\mathbf{V}}:=\frac{1}{\varepsilon^3}\int_0^{\varepsilon^3}\!\!\!\!d^3x\sqrt{p^1p^2p^3}=\sqrt{p^1p^2p^3}=\varepsilon^3\sqrt{\bar{q}}\,.
\end{align}
The square root of operators after quantization is going to be derived by the expansion of the radicand around the coherent state, resulting in the expression
\begin{align}
\label{volume_expansion}
\hat{\bar{\mathbf{V}}}^n
=\varepsilon^{3n}\big(\langle\hat{\bar{q}}\rangle\big)^{\@\frac{n}{2}}
\sum_{k=0}^{\infty}\binom{n/2}{k}\bigg(\frac{\hat{\bar{q}}-\langle\hat{\bar{q}}\rangle}{\langle\hat{\bar{q}}\rangle}\bigg)^{\!\!k},
\end{align}
where the expectation value of the $\hat{\bar{q}}$ operator is
\begin{align}
\label{volume_coherent}
\langle\hat{\bar{q}}\rangle=\kbar^3\.\bar{m}^1\bar{m}^2\bar{m}^3\,.
\end{align}

The reduced holonomy in \eqref{gr_coupling_matter} leads to the states modifications indicated in \eqref{red_action_c}. The action of the volume operator in \eqref{volume_expansion} on these modified states can be expressed in the form of a power series,
\begin{align}
\label{volume_shifted}
\begin{split}
\hat{\bar{\mathbf{V}}}^n\bigg|\bar{m}^{(i)}_v\pm\frac{1}{2}\bigg>
=&\;\big(\kbar^3\.\bar{m}^1_v\.\bar{m}^2_v\.\bar{m}^3_v\big)^{\@\frac{n}{2}}\Bigg[
1
\pm\frac{n}{4}\frac{1}{\bar{m}^{(i)}_v\@}
+\frac{n(n\@-\@2)}{2^5}\frac{1}{\big(\bar{m}^{(i)}_v\big)^{\@2}\@}
\\
&\;\pm\frac{n(n\@-\@2)(n\@-\@4)}{2^7\@\cdot\@3}\frac{1}{\big(\bar{m}^{(i)}_v\big)^{\@3}\@}
+\mathcal{O}\Bigg(\frac{1}{\big(\bar{m}^{(i)}_v\big)^{\@4}\@}\Bigg)\Bigg]
\bigg|\bar{m}^{(i)}_v\pm\frac{1}{2}\bigg>\,.
\end{split}
\end{align}
Here, the large quantum numbers assumption was needed. Then, the action of the quantum GR corrections generating operator is easily calculable and reads
\begin{align}
\label{gr_matter}
\!\text{tr}\bigg(\@\tau^{(i)}\hat{h}_{\@(j\@)}^{-1}\Big[\hat{\mathbf{V}}^n,\hat{h}_{\@(j\@)}^{\mathstrut}\Big]\@\bigg)
\@\bigotimes_k^3\!\Big|\bar{m}^{(k)}_v\@\Big>
=
\frac{\i\.n}{4\bar{m}^{(i)}_v\!}\big(\kbar^3\.\bar{m}^1_v\.\bar{m}^2_v\.\bar{m}^3_v\big)^{\@\frac{n}{2}}\delta^{(i)}_{(j)}
\Bigg[
1
+\frac{\@n^2\@-6n\@+8\@}{2^5\@\cdot\@3}\frac{1}{\!\big(\bar{m}^{(i)}_v\big)^{\@2}\!}
+\mathcal{O}\@\Bigg(\@\frac{1}{\@\big(\bar{m}^{(i)}_v\big)^{\@4}\!}\!\Bigg)\Bigg]
\@\bigotimes_k^3\!\Big|\bar{m}^{(k)}_v\@\Big>\,.
\!
\end{align}
Consequently, the values of the dimensionless corrections in \eqref{corrections_explicit_vector}, \eqref{corrections_explicit_mom}, and \eqref{corrections_explicit_der} are
\begin{align}
\label{correction_vector}
\delta^{(\underline{A})}_{\@(i)\mathstrut\!}
=&\;\frac{7}{2^6}\frac{1}{\@\big(\bar{m}^{(i)}_v\big)^{\@2}\@}
\to7\pi^2\gamma^2\frac{l_P^4}{\varepsilon^4}\frac{q_{(\@i\@)\@(\@i\@)}}{|\bar{q}|}\,,
\\
\label{correction_mom}
\delta^{(\varphi)}_{\@_\text{mom}\mathstrut\!}
=&\;\frac{7}{2^6}\sum_i^3\frac{1}{\@\big(\bar{m}^{(i)}_v\big)^{\@2}\@}
=\sum_i^3\delta^{(\underline{A})}_{\@(i)\mathstrut\!}\,,
\intertext{and}
\label{correction_der}
\delta^{(\varphi)}_{\@(i)\._\text{der}\mathstrut\!}
=&\;\frac{65}{2^8\@\cdot\@3}\sum_{j\neq i}\frac{1}{\@\big(\bar{m}^{(j)}_v\big)^{\@2}\@}
=\frac{65}{84}\sum_{j\neq i}\delta^{(\underline{A})}_{\@(j)\mathstrut\!}\,,
\end{align}
respectively.

For the analysis in this article, only the precise structure of corrections is needed. However, the reader interested in the explicit values can easily derive each correction by using the correspondence principle in \eqref{correspondence} --- as demonstrated on  the right-hand side of formula \eqref{correction_vector}. The ${l_P/\varepsilon}$ ratio must be regularized by a cutoff on the value of the regulator, which has to be consistent with the large spin approximation in \eqref{volume_shifted}. The condition ${\varepsilon^2>|\gamma|\.l_P^2}$ is acceptable, but it should be replaced with ${\varepsilon^2\ggg\gamma\.l_P^2}$. Otherwise, keeping the trigonometric form of the reduced holonomy in \eqref{gr_coupling_matter} has no sense and the approximation ${\sin\big(\hat{c}_{(\@a\@)}^{\.i}/2\big)}\simeq{\hat{c}_{(\@a\@)}^{\.i}/2}$, ${\cos\big(\hat{c}_{(\@a\@)}^{\.i}/2\big)}\simeq1$ is indistinguishable from that form. This occurs for instance by using the so-called area gap, $\varepsilon^2\approx{25\.\gamma\.l_P^2}$, \textit{cf.} \cite{Ashtekar:2011ni}. This cutoff imposed on the phase space-reduced CQGR leads to the domination of the inverse volume corrections over any other quantum corrections and the HCO becomes almost exactly an eigenvector of the basis states.

Finally, one can test BI of the cosmologically reduced CQGR. By applying the phase space reduction, all the gauge symmetries of the theory became restricted to their reduced versions, but not violated or modified. Therefore by the inspection of the reduced diffeomorphism transformations of the CQGR semiclassical limit, general covariance can be verified.

The main vector field observables are the $\hat{\underline{E}}^a$ and $\hat{\underline{B}}^a$ operators. Their expectation values are the electric vector field density and the magnetic pseudovector field density. They are the weight $1$ physical quantities that are the measurable modes of an electromagnetic wave. Another observable is the HCO. Its expectation value is the weight $1$ scalar density that represents the energy density. The difference between the energy densities related to different spacetime points is also an explicitly measurable quantity. The semiclassical limit of the related HCO reads
\begin{align}
\label{expectation_vector}
\big<\hat{H}^{(\underline{A})}\big>
=
\sum_i^3\big<\hat{H}^{(\underline{A})}_{\@(i)}\big>
=
\sum_i^3\big<\hat{H}^{(\underline{A})}_{\@(i)}\big>^{\@(\underline{A})}\Big(
1+\delta^{(\underline{A})}_{\@(i)\mathstrut\!}
\Big)\,.
\end{align}
It is clear that the reduced diffeomorphism transformations describable by the metric tensor contraction are the same in the electric and magnetic elements of the scalar constraint. Moreover, the corrections in \eqref{expectation_vector} are preserved concerning the temporal diffeomorphism transformation, \textit{i.e.} these corrections are independent of dynamics --- see \eqref{corrections_Hamiltonian_vector}. Therefore, although the quantum GR corrections apparently change the contraction with the metric tensor, the relative diffeomorphism symmetry along all the reduced directions is not modified.

Considering then the quantum system of the vector field and gravity, one could wish to be able to restore the explicit classical form of the general covariance specifying spatial metric tensor. This would allow one to apply the methods of QFT on curved spacetime to the expectation values of the matter degrees of freedom in an effective model. This effective procedure would be possible by assuming that the Hamiltonian density of the whole system equals zero. Then, the effective `covarianization' method can be defined as the multiplication of all the scalar constraint elements by the inverse of $\Big(1+\delta^{(\underline{A})}_{\@(i)\mathstrut\!}\Big)$. In this way, the GR corrections will be moved to the Hamiltonian gravitational contribution. This contribution will effectively represent the gravitational sector of the scalar constraint coupled with the standard QFT representation of the electromagnetic field on classical curved spacetime.

In the case of the scalar field, the situation is completely different. It is worth emphasizing that the variable $\pi$ is a weight $1$ pseudoscalar density, but $\partial_a\varphi$ and $\varphi$ are a one-form and a scalar, respectively. This explains why the semiclassical limit of the HCO has different GR corrections for each element,
\begin{align}
\label{expectation_scalar}
\begin{split}
\big<\hat{H}^{(\varphi)}\big>
=&\,
\sum_i^3\bigg[
\Big<\hat{H}^{(\varphi)}_{\@(i)\._\text{mom}\mathstrut\!}\Big>
+\Big<\hat{H}^{(\varphi)}_{\@(i)\._\text{der}\mathstrut\!}\Big>
+\Big<\hat{H}^{(\varphi)}_{\@(i)\._\text{pot}\mathstrut\!}\Big>
\bigg]
=
\sum_i^3\bigg[
\Big<\hat{H}^{(\varphi)}_{\@(i)\._\text{mom}\mathstrut\!}\Big>^{\@\@(\varphi)}
\Big(1+2\delta^{(\underline{A})}_{\@(i)\mathstrut\!}\Big)
\\
&\;
+\Big<\hat{H}^{(\varphi)}_{\@(i)\._\text{der}\mathstrut\!}\Big>^{\@\@(\varphi)}
\bigg(1-\delta^{(\underline{A})}_{\@(i)\mathstrut\!}+\frac{65}{84}\sum_{j\neq i}\delta^{(\underline{A})}_{\@(j)\mathstrut\!}\bigg)
+\Big<\hat{H}^{(\varphi)}_{\@(i)\._\text{pot}\mathstrut\!}\Big>^{\@\@(\varphi)}
\Big(1-\delta^{(\underline{A})}_{\@(i)\mathstrut\!}\Big)
\bigg]
\Big(1+\delta^{(\underline{A})}_{\@(i)\mathstrut\!}\Big)
+\mathcal{O}\bigg(\Big(\delta^{(\underline{A})}_{\@(i)\mathstrut\!}\Big)^{\@\@2}\bigg),
\end{split}
\end{align}
where
$\Big<\hat{H}^{(\varphi)}_{\@1\._\text{mom}\mathstrut\!}\Big>
=\Big<\hat{H}^{(\varphi)}_{\@2\._\text{mom}\mathstrut\!}\Big>
=\Big<\hat{H}^{(\varphi)}_{\@3\._\text{mom}\mathstrut\!}\Big>
=\frac{1}{3}\Big<\hat{H}^{(\varphi)}_{\@_\text{mom}\mathstrut\!}\Big>$
and
$\Big<\hat{H}^{(\varphi)}_{\@1\._\text{pot}\mathstrut\!}\Big>
=\Big<\hat{H}^{(\varphi)}_{\@2\._\text{pot}\mathstrut\!}\Big>
=\Big<\hat{H}^{(\varphi)}_{\@3\._\text{pot}\mathstrut\!}\Big>
=\frac{1}{3}\Big<\hat{H}^{(\varphi)}_{\@_\text{pot}\mathstrut\!}\Big>$.
Therefore the relative diffeomorphism symmetry is not preserved at the level of corrections. These GR corrections are the semiclassical predictions, hence they are potentially measurable quantities. This asymmetry indicates the background dependence of the result and thus breaks general covariance. Even by neglecting the self-interaction terms, the relative local diffeomorphism symmetry of the momentum and derivative sectors is not equivalent. Consequently, the effective covarianization procedure is also not applicable to the expression in \eqref{expectation_scalar}.

The last result demonstrates that the node representation leads to the background-dependent structure of GR corrections. The simplest resolution of this problem is to apply the isotropic vector field representation to describe the quantity that classically is expressed by the scalar field.

By assuming this representation, the effective Hamiltonian of the bosonic system on a cuboidal lattice can be written as the following sum:
\begin{align}
\label{cosmo_scalar}
\bar{H}=\sum_i^3
\Big(\bar{H}^{(\text{gr}\@)}_{\@(i)}+\bar{H}^{(\und{A})}_{\@(i)}+\bar{H}^{(\varphi)}_{\@(i)}\Big)
=:\sum_i^3\bar{H}_{\@(i)}\,.
\end{align}
One can assume that the self-interacting scalar field is represented by the isotropic Proca Hamiltonian. Then, by fixing the total energy to zero by setting $H=0$, the effective covarianization method will be the removal of the GR corrections in the procedure defined by
\begin{align}
\label{cosmo_covarianization}
\bar{H}_{\@(i)}\,\stackrel{\!\text{\tiny covar.}}{\longrightarrow}\,\bar{H}_{\@(i)}\Big(1+\delta^{(\underline{A})}_{\@(i)\mathstrut\!}\Big)^{\@\@-1}.
\end{align}
As a result, the whole free matter sector becomes corrections independent. However, the mass term and any other potential contribution becomes shifted down by the factor $\big(1-\sum_i^3\delta^{(\underline{A})}_{\@(i)\mathstrut\!}\big)$. This is an interesting prediction, however, it is not a fundamental theory result. Therefore, as in the case of the effective models recalled in Sec.~\ref{II.5}, one should not consider this outcome as a prediction, for instance, of the inflaton field's real mass loss. It could be only used to effectively describe this phenomenon if it would be observed.

An even more interesting observation concerns the inclusion of the fermionic sector. The scalar constraint of the system that describes all fundamental interactions in the cosmologically reduced framework can be effectively expressed by
\begin{align}
\begin{split}
\label{general_scalar}
H=&\,\int_{\Sigma_t}\!\!\!\!d^3x\,N
\bigg[
\frac{1}{\kappa\sqrt{q}}\Big(\big(F^i_{cd}-(\gamma^2+1)\epsilon_{ilm}K^l_cK^m_d\big)\epsilon^{ijk}E_j^aE_k^b\Big)
+
\frac{\mathsf{g}_{\!\und{A}}^2\!}{2\sqrt{q}}\.q_{cd}\big(\und{E}^a_I\und{E}^b_I+\und{B}^a_I\und{B}^b_I\big)
\\
&\;
+\frac{2\sqrt{q}}{3\kappa}\Lambda\.q_{cd}q^{ab}
+\mathcal{H}^{(\varphi)}{}_{cd}\.q^{ab}
+\mathcal{H}^{(\psi)}{}_{cd}^{ab}
\bigg]\delta^c_a\delta^d_b
=:H_{cd}^{ab}
\,\delta^c_a\delta^d_b\,.
\end{split}
\end{align}
Here, all the matter sector is assumed to be smeared along the links of the cuboidal lattice. The torsional contribution from the fermionic sector is assumed to by given by the procedure in \cite{Bojowald:2007nu,Bojowald:2007pc} and the regularization of the Dirac contribution follows either the method in \cite{Thiemann:1997rt,Thiemann:2007zz} or the one in \cite{Bojowald:2007nu,Bojowald:2007pc} adjusted to the links smearing of the fermionic field. Then the analog of \eqref{H_vect} and \eqref{H_scal} is the following Hamiltonian constraint:
\begin{align}
\label{scalar_fermion}
H^{(\psi)}=\!\int_{\Sigma_t}\!\!\!\!d^3x\,\frac{N}{\sqrt{q}}
\bigg[
\.\epsilon_{ijk}\.\epsilon^{abe}
\,\text{tr}\bigg(
\tau^i\frac{1}{\varepsilon}h_c^{-1}\@(x)\Big\{\mathbf{V}^{\frac{1}{2}}\@(x),h_c^{\mathstrut}\@(x)\Big\}
\bigg)
\,\text{tr}\bigg(
\tau^j\frac{1}{\varepsilon}h_d^{-1}\@(x)\Big\{\mathbf{V}^{\frac{1}{2}}\@(x),h_d^{\mathstrut}\@(x)\Big\}
\bigg)
(\text{fermionic})_e^k
\bigg]
\delta^c_a\delta^d_b\,.
\end{align}
Here, `$(\text{fermionic})$' denotes the Dirac field's degrees of freedom. Consequently, the related quantum GR corrections take analogous form to \eqref{expectation_vector}, but they are antisymmetric. Conversely, the structure of the bosonic fields Hamiltonian has the form $H_{cd}^{ab}\,\delta^c_a\delta^d_b$, hence it is symmetric in the pairs of spatial-internal indices (the external and internal directions are indistinguishable after the reduction). This breaks the BI of the system, however, this violation occurred in the expected manner. One can anticipate that in the antisymmetric sector of CQGR, the corresponding  diffeomorphism invariance will be correctly preserved. To complete this conjecture, it is worth it to mention that the related covarianization will be given by the expression
\begin{align}
\label{general_covarianization}
H_{cd}^{ab}(v)\,\stackrel{\!\text{\tiny covar.}}{\longrightarrow}\,H_{cd}^{ab}(v)
\bigg(1+\frac{1}{2}\delta^{(\underline{A})}_{\@(a)\mathstrut\!}\@(v)\bigg)^{\!\!-1}
\bigg(1+\frac{1}{2}\delta^{(\underline{A})}_{\@(b)\mathstrut\!}\@(v)\bigg)^{\!\!-1}.
\end{align}


\section{Conclusions}\label{V}

\noindent
This article revealed the problems with the accurate implementation of general covariance in the matter sector of CQGR, where the theory is assumed to be constructed by using the LQG's framework. By general covariance, the BI condition originally postulated by Einstein was considered. This quantity together with SE, called also the equivalence principle, forms the general principle of relativity.

The BI violation was a consequence of using inconsistent regularization methods. This inconsistency was regarding the local spatial diffeomorphism symmetry breakdown in the continuous to discrete transition of the multifield system. Then, the lack of general covariance was revealed in the structure of the semiclassical corrections of the cosmologically reduced CQGR.

In the LQG's framework, the symmetry of the canonical fields lattice smearing is the symmetry of the links of this lattice. The links structure specifies the discrete diffeomorphism transformations directions distribution. Therefore, it is not surprising that by using the locally diffeomorphisms breaking representation of a field located at nodes, the general covariance of the system is violated. It should be emphasized that the diffeomorphism symmetry becomes locally broken in the following series of steps. First the propagating gravitational degrees of freedom are smeared by using the holonomy-flux representation, where the relation in \eqref{E_trick} is assumed. Next, the phase space reduction, which preserves all the reduced symmetries, is implemented. Then, the theory is quantized and the semiclassical limit is derived on the Gaussian states that are picked at the momenta (or volume) eigenvalues. Finally, by the correspondence principle, the original metric structure is restored and its asymmetry in the scalar field Hamiltonian elements is revealed. What needs to be added to this list is the fact that the scalar field degrees of freedom were lattice regularized at nodes, conversely to all of the other variables, which were smeared accordingly to the links' structure. Moreover, all but the first step were exact, however, this step considered only the essential techniques of LQG. Furthermore, the approximations in this step (before quantization) were reproducing the original continuous formulation of the theory exactly, by taking the limit $\varepsilon\to0$. Anything that could be questioned in the analysis in this article concerns the methods of LQG. The phase space reduction was implemented in the standard manner \cite{Henneaux:1992ig} in which the Dirac brackets take the form of the Poisson ones, \textit{cf.} \cite{Bilski:2019tji}.

Concluding, the following no-go theorem concerning the lattice regularization in the framework of LQG can be formulated. {\sl Let a model of quantum general relativity be considered, where the loop quantum gravitational techniques are used to regularize and quantize gravitational degrees of freedom. By assuming the systems-equivalent description and background-independent predictions of this model, the lattice regularization of matter minimally coupled to gravity is restricted. The matter variables selected for the lattice smearing should be represented by vector densities to ensure that all the coupled gravitational degrees of freedom are written in an appropriate form. Moreover, this representation allows to express the matter degrees of freedom on the lattice in terms of the holonomy-flux formalism, which is also the representation of the gravitational variables. By choosing a nodes smearing, the general covariance of the theory predictions will be violated.} Furthermore, it is worth it to add that in the case of the properly lattice-regularized electromagnetic field, the smeared variables are the ones that have the explicit and real physical meaning. They are the electric and magnetic fields.

One more comment is worth it to be added at the end. In this article it was not certainly demonstrated that the fermionic matter must be lattice-regularized accordingly to the aforementioned theorem. However, so suggesting indications were found. Therefore, it is probable that the fermionic variables proposed in the context of LQG (represented by the Grassmann-valued scalar half-densities) \cite{Thiemann:2007zz,Thiemann:1997rt,Thiemann:1997rq,Bojowald:2007nu} should be replaced by appropriate vector half-densities. The weight $1/2$ would reflect the fermionic otherness from the weight $1$ of the vector representations of bosons.


\section*{Acknowledgements}

\noindent
The author thanks  Suddhasattwa Brahma for discussions.
This research was partially supported by the National Natural Science Foundation of China Grant Nos. 11675145 and 11975203.




\begin{thebibliography}{99}

\bibitem{Thiemann:1996aw}
T.~Thiemann,
Class.\ Quant.\ Grav.\  {\bf 15}, 839 (1998)
[gr-qc/9606089].

\bibitem{Thiemann:2007zz}
T.~Thiemann,
Cambridge, UK: Cambridge Univ. Pr. (2007).

\bibitem{Ashtekar:1986yd}
A.~Ashtekar,
Phys.\ Rev.\ Lett.\  {\bf 57}, 2244 (1986).

\bibitem{Yang:1954ek} 
C.~N.~Yang and R.~L.~Mills,
Phys.\ Rev.\  {\bf 96}, 191 (1954).
doi:10.1103/PhysRev.96.191

\bibitem{Wilson:1974sk} 
K.~G.~Wilson,
Phys.\ Rev.\ D {\bf 10}, 2445 (1974).

\bibitem{Einstein:1916vd}
A.~Einstein,
Annalen Phys. \textbf{354}, no.7, 769-822 (1916)
doi:10.1002/andp.200590044

\bibitem{Einstein:1907} 
A.~Einstein,
Jahrb Radioaktivit\"{a}t Elektronik {\bf 4}, 411-462 (1907).

\bibitem{Einstein:1911vc}
A.~Einstein,
Annalen Phys. \textbf{340}, 898-908 (1911)
doi:10.1002/andp.200590033

\bibitem{Einstein:1907iag}
A.~Einstein,
Annalen Phys. \textbf{328}, no.7, 371-384 (1907)
doi:10.1002/andp.19073280713

\bibitem{DeWitt:1967yk} 
B.~S.~DeWitt,
Phys.\ Rev.\  {\bf 160}, 1113 (1967).
doi:10.1103/PhysRev.160.1113

\bibitem{Ashtekar:2009vc} 
A.~Ashtekar and E.~Wilson-Ewing,
Phys.\ Rev.\ D {\bf 79}, 083535 (2009)
doi:10.1103/PhysRevD.79.083535
[arXiv:0903.3397 [gr-qc]].

\bibitem{Ashtekar:2011ni}
A.~Ashtekar and P.~Singh,
Class. Quant. Grav. \textbf{28}, 213001 (2011)
doi:10.1088/0264-9381/28/21/213001
[arXiv:1108.0893 [gr-qc]].

\bibitem{Thiemann:1997rt}
T.~Thiemann,
Class.\ Quant.\ Grav.\  {\bf 15}, 1281 (1998)
[gr-qc/9705019].

\bibitem{Bilski:2015dra}
J.~Bilski, E.~Alesci and F.~Cianfrani,
Phys.\ Rev.\ D {\bf 92}, no. 12, 124029 (2015)
doi:10.1103/PhysRevD.92.124029
[arXiv:1506.08579 [gr-qc]].

\bibitem{Bilski:2019tji}
J.~Bilski and A.~Marcianò,
Phys. Rev. D \textbf{101}, no.6, 066026 (2020)
doi:10.1103/PhysRevD.101.066026
[arXiv:1905.00001 [gr-qc]].

\bibitem{Bilski:2016pib} 
J.~Bilski, E.~Alesci, F.~Cianfrani, P.~Don\`a and A.~Marcian\`o,
Phys.\ Rev.\ D {\bf 95}, no. 10, 104048 (2017)
doi:10.1103/PhysRevD.95.104048
[arXiv:1612.00324 [gr-qc]].

\bibitem{Arnowitt:1960es} 
R.~L.~Arnowitt, S.~Deser and C.~W.~Misner,
Phys.\ Rev.\  {\bf 117}, 1595 (1960).
doi:10.1103/PhysRev.117.1595

\bibitem{Arnowitt:1962hi}
R.~L.~Arnowitt, S.~Deser and C.~W.~Misner,
Gen.\ Rel.\ Grav.\  {\bf 40}, 1997 (2008)
[gr-qc/0405109].

\bibitem{Ashtekar:2002sn} 
A.~Ashtekar, S.~Fairhurst and J.~L.~Willis,
Class.\ Quant.\ Grav.\  {\bf 20}, 1031 (2003)
doi:10.1088/0264-9381/20/6/302
[gr-qc/0207106].

\bibitem{Bilski:2020poi}
J.~Bilski,
[arXiv:2012.14441 [gr-qc]].

\bibitem{Bilski:2021_LCC}
J.~Bilski and A.~Wang,
[arXiv:2101.02223 [gr-qc]].

\bibitem{Barbero:1994ap} 
J.~F.~Barbero G.,
Phys.\ Rev.\ D {\bf 51}, 5507 (1995)
doi:10.1103/PhysRevD.51.5507
[gr-qc/9410014].

\bibitem{Bilski:2020xfq}
J.~Bilski,
[arXiv:2012.10465 [gr-qc]].

\bibitem{Thiemann:1997rq}
T.~Thiemann,
Class.\ Quant.\ Grav.\  {\bf 15}, 1487 (1998)
[gr-qc/9705021].

\bibitem{Ashtekar:2002vh} 
A.~Ashtekar, J.~Lewandowski and H.~Sahlmann,
Class.\ Quant.\ Grav.\  {\bf 20}, L11 (2003)
[gr-qc/0211012].

\bibitem{Kaminski:2005nc} 
W.~Kaminski, J.~Lewandowski and M.~Bobienski,
Class.\ Quant.\ Grav.\  {\bf 23}, 2761 (2006)
[gr-qc/0508091].

\bibitem{Kaminski:2006ta} 
W.~Kaminski, J.~Lewandowski and A.~Okolow,
Class.\ Quant.\ Grav.\  {\bf 23}, 5547 (2006)

\bibitem{Hossain:2010wy}
G.~M.~Hossain, V.~Husain and S.~S.~Seahra,
Class. Quant. Grav. \textbf{27}, 165013 (2010)
doi:10.1088/0264-9381/27/16/165013
[arXiv:1003.2207 [gr-qc]].

\bibitem{Wilson-Ewing:2016yan} 
E.~Wilson-Ewing,
Comptes Rendus Physique {\bf 18}, 207 (2017)
doi:10.1016/j.crhy.2017.02.004
[arXiv:1612.04551 [gr-qc]].

\bibitem{Bojowald:2012xy} 
M.~Bojowald,
Class.\ Quant.\ Grav.\  {\bf 29}, 213001 (2012)
doi:10.1088/0264-9381/29/21/213001
[arXiv:1209.3403 [gr-qc]].

\bibitem{Barrau:2013ula} 
A.~Barrau, T.~Cailleteau, J.~Grain and J.~Mielczarek,
Class.\ Quant.\ Grav.\  {\bf 31}, 053001 (2014)
doi:10.1088/0264-9381/31/5/053001
[arXiv:1309.6896 [gr-qc]].

\bibitem{Barrau:2014maa}
A.~Barrau, M.~Bojowald, G.~Calcagni, J.~Grain and M.~Kagan,
JCAP \textbf{05}, 051 (2015)
doi:10.1088/1475-7516/2015/05/051
[arXiv:1404.1018 [gr-qc]].
  
\bibitem{Ashtekar:2003hd} 
A.~Ashtekar, M.~Bojowald and J.~Lewandowski,
Adv.\ Theor.\ Math.\ Phys.\  {\bf 7}, no. 2, 233 (2003)
doi:10.4310/ATMP.2003.v7.n2.a2
[gr-qc/0304074].

\bibitem{Bojowald:2008zzb} 
M.~Bojowald,
Living Rev.\ Rel.\  {\bf 11}, 4 (2008).

\bibitem{Ashtekar:2009mb} 
A.~Ashtekar, W.~Kaminski and J.~Lewandowski,
Phys.\ Rev.\ D {\bf 79}, 064030 (2009)
doi:10.1103/PhysRevD.79.064030
[arXiv:0901.0933 [gr-qc]].

\bibitem{Agullo:2012fc} 
I.~Agullo, A.~Ashtekar and W.~Nelson,
Phys.\ Rev.\ D {\bf 87}, no. 4, 043507 (2013)
doi:10.1103/PhysRevD.87.043507
[arXiv:1211.1354 [gr-qc]].

\bibitem{Lewandowski:2017cvz} 
J.~Lewandowski, M.~Nouri-Zonoz, A.~Parvizi and Y.~Tavakoli,
Phys.\ Rev.\ D {\bf 96}, no. 10, 106007 (2017)
doi:10.1103/PhysRevD.96.106007
[arXiv:1709.04730 [gr-qc]].

\bibitem{FernandezMendez:2012vi} 
M.~Fernandez-Mendez, G.~A.~Mena Marugan and J.~Olmedo,
Phys.\ Rev.\ D {\bf 86}, 024003 (2012)
doi:10.1103/PhysRevD.86.024003
[arXiv:1205.1917 [gr-qc]].

\bibitem{Gomar:2014faa}
L.~C.~Gomar, M.~Fern\'andez-M\'endez, G.~A.~M.~Marug\'an and J.~Olmedo,
Phys. Rev. D \textbf{90}, no.6, 064015 (2014)
doi:10.1103/PhysRevD.90.064015
[arXiv:1407.0998 [gr-qc]].

\bibitem{Wilson-Ewing:2015sfx} 
E.~Wilson-Ewing,
Int.\ J.\ Mod.\ Phys.\ D {\bf 25}, no. 08, 1642002 (2016)
doi:10.1142/S0218271816420025
[arXiv:1512.05743 [gr-qc]].

\bibitem{Bojowald:2020xlw} 
M.~Bojowald,
arXiv:2002.04986 [gr-qc].
  
\bibitem{Bojowald:2020wuc}
M.~Bojowald,
Universe \textbf{6}, no.3, 36 (2020)
doi:10.3390/universe6030036
[arXiv:2002.05703 [gr-qc]].

\bibitem{Bojowald:2020unm}
M.~Bojowald,
Phys. Rev. D \textbf{102}, no.4, 046006 (2020)
doi:10.1103/PhysRevD.102.046006
[arXiv:2007.16066 [gr-qc]].

\bibitem{Thiemann:1996av}
T.~Thiemann,
Class. Quant. Grav. \textbf{15}, 875-905 (1998)
doi:10.1088/0264-9381/15/4/012
[arXiv:gr-qc/9606090 [gr-qc]].

\bibitem{Thiemann:1997rv}
T.~Thiemann,
Class. Quant. Grav. \textbf{15}, 1207-1247 (1998)
doi:10.1088/0264-9381/15/5/010
[arXiv:gr-qc/9705017 [gr-qc]].

\bibitem{Alesci:2013xd}
E.~Alesci and F.~Cianfrani,
Phys.\ Rev.\ D {\bf 87}, no. 8, 083521 (2013)
[arXiv:1301.2245 [gr-qc]].

\bibitem{Alesci:2014uha} 
E.~Alesci and F.~Cianfrani,
Phys.\ Rev.\ D {\bf 90}, no. 2, 024006 (2014)
doi:10.1103/PhysRevD.90.024006
[arXiv:1402.3155 [gr-qc]].

\bibitem{Dapor:2017gdk} 
A.~Dapor and K.~Liegener,
Class.\ Quant.\ Grav.\  {\bf 35}, no. 13, 135011 (2018)
doi:10.1088/1361-6382/aac4ba
[arXiv:1710.04015 [gr-qc]].

\bibitem{Yang:2009fp} 
J.~Yang, Y.~Ding and Y.~Ma,
Phys.\ Lett.\ B {\bf 682}, 1 (2009)
doi:10.1016/j.physletb.2009.10.072
[arXiv:0904.4379 [gr-qc]].

\bibitem{Bojowald:2001vw} 
M.~Bojowald,
Phys.\ Rev.\ D {\bf 64}, 084018 (2001)
doi:10.1103/PhysRevD.64.084018
[gr-qc/0105067].

\bibitem{Grain:2009cj} 
J.~Grain, A.~Barrau, and A.~Gorecki,
Phys.\ Rev.\ D {\bf 79}, 084015 (2009)
doi:10.1103/PhysRevD.79.084015
[arXiv:0902.3605 [gr-qc]].

\bibitem{Flori:2008nw} 
C.~Flori and T.~Thiemann,
arXiv:0812.1537 [gr-qc].

\bibitem{Heisenberg:1929xj}
W.~Heisenberg and W.~Pauli,
Z. Phys. \textbf{56}, 1-61 (1929)
doi:10.1007/BF01340129.

\bibitem{Thiemann:2000bw}
T.~Thiemann,
Class. Quant. Grav. \textbf{18}, 2025-2064 (2001)
doi:10.1088/0264-9381/18/11/304
[arXiv:hep-th/0005233 [hep-th]].

\bibitem{Thiemann:2000ca}
T.~Thiemann and O.~Winkler,
Class. Quant. Grav. \textbf{18}, 2561-2636 (2001)
doi:10.1088/0264-9381/18/14/301
[arXiv:hep-th/0005237 [hep-th]].

\bibitem{Thiemann:2000bx}
T.~Thiemann and O.~Winkler,
Class. Quant. Grav. \textbf{18}, 4629-4682 (2001)
doi:10.1088/0264-9381/18/21/315
[arXiv:hep-th/0005234 [hep-th]].

\bibitem{Dittrich:2014ala} 
B.~Dittrich,
doi:10.1142/9789813220003\_0006
[arXiv:1409.1450 [gr-qc]].

\bibitem{Bronstein_cube}
M.~P.~Bronstein,
Uspekhi Astronomicheskikh Nauk. Sbornik, No. {\bf 3}, Moscow, USSR: ONTI, 3-30 (1933).

\bibitem{Birrell:1982ix} 
N.~D.~Birrell and P.~C.~W.~Davies,
doi:10.1017/CBO9780511622632

\bibitem{Guth:1980zm} 
A.~H.~Guth,
Phys.\ Rev.\ D {\bf 23}, 347 (1981)
[Adv.\ Ser.\ Astrophys.\ Cosmol.\  {\bf 3}, 139 (1987)].
doi:10.1103/PhysRevD.23.347

\bibitem{Bojowald:1999tr} 
M.~Bojowald,
Class.\ Quant.\ Grav.\  {\bf 17}, 1489 (2000)
doi:10.1088/0264-9381/17/6/312
[gr-qc/9910103].

\bibitem{Bojowald:2001xe} 
M.~Bojowald,
Phys.\ Rev.\ Lett.\  {\bf 86}, 5227 (2001)
doi:10.1103/PhysRevLett.86.5227
[gr-qc/0102069].

\bibitem{Bojowald:2007nu} 
M.~Bojowald and R.~Das,
Phys.\ Rev.\ D {\bf 78}, 064009 (2008)
doi:10.1103/PhysRevD.78.064009
[arXiv:0710.5722 [gr-qc]].

\bibitem{Bojowald:2007pc}
M.~Bojowald, R.~Das and R.~J.~Scherrer,
Phys. Rev. D \textbf{77}, 084003 (2008)
doi:10.1103/PhysRevD.77.084003
[arXiv:0710.5734 [astro-ph]].

\bibitem{Henneaux:1992ig} 
M.~Henneaux and C.~Teitelboim,
Princeton, USA: Univ. Pr. (1992).

\end{thebibliography}
\end{document}